\documentclass[11pt]{amsart}

\calclayout

\makeatletter
\def\subsection{\@startsection{subsection}{2}%
  \z@{.5\linespacing\@plus.7\linespacing}{.5\baselineskip}%
  {\normalfont\centering\bfseries}%
}

\def\subsubsection{\@startsection{subsubsection}{3}%
  \z@{.5\linespacing\@plus.7\linespacing}{.5\baselineskip}%
  {\normalfont\centering\itshape}%
}
\makeatother

\usepackage{graphicx}
\usepackage{newtxtext}
\usepackage{newtxmath}
\usepackage{natbib}
\usepackage{hyperref}
\usepackage{comment}
\usepackage{amsmath}
\DeclareRobustCommand*\cal{\mathcal}
 \usepackage{ulem}
\hypersetup{
    colorlinks = true,
    urlcolor   = blue,
    citecolor  = black,
}

\newcommand{\RomanNumeralCaps}[1]
\linenumbers

\title{Inverse Transfer and Coherence in Rotating Stratified Flow with Clouds and Phase Transitions}

\author{Yeyu Zhang\textsuperscript{1,*}, 
        Yingshuo Peng\textsuperscript{1}, 
        Leslie M. Smith\textsuperscript{2,3}}

\begin{document}
\maketitle
\vspace{-1em}
\begin{center}
{\small
\textsuperscript{1} School of Mathematics, Shanghai University of Finance and Economics, Shanghai 200433, PR China\\
\textsuperscript{2} Department of Mathematics, University of Wisconsin–Madison, WI 53706, USA\\
\textsuperscript{3} Department of Atmospheric and Oceanic Sciences, University of Wisconsin–Madison, WI 53706, USA\\
\vspace{0.5em}
\textsuperscript{*} \texttt{zhangyeyu@mail.shufe.edu.cn}
}
\end{center}

\begin{abstract}
Inverse energy transfer to large-scale coherent structures in idealized models of geophysical flows has been 
% a subject 
of interest for 
 over four decades. Extensive knowledge exists regarding inverse %energy 
transfer in rotating and stratified dry dynamics, characterized by the Rossby number and a single dry Froude number. 
The current study includes effects of water and phase changes, 
with dynamics characterized by the Rossby number and two Froude numbers for unsaturated and saturated environments. Using numerical computations with random forcing,  inverse energy transfer 
% of energy
is examined for a model with a Boussinesq dynamical core, incorporating water vapor and liquid water in the limit of asymptotically-fast cloud microphysics. Besides kinetic energy, 
% the
total energy includes buoyant potential energies from each phase, 
% as well as 
 and latent moist energy responsible for potential energy transfer at phase boundaries. The rotation and stratification terms are large and comparable, such that the dry version of the evolution equations is dominated by inverse transfer of pseudo potential vorticity($PV$). 
 
% It is found that 
%Upscale transfer of energy persists 
% in the presence of 
%with phase changes for cloud fraction of $59\%$. 
For fixed Rossby and dry (unsaturated) Froude numbers, compared to dry dynamics, there is a reduction in energy transfer rate, associated with the larger Froude number of saturated regions. 
%Phase changes and latent heat exchange increase the ratio of potential and kinetic energies and decrease overall energy transfer to large scales. However, the upscale transfer persists and is influenced by nonlinear waves at lowest order resulting from nonlinear buoyancy near phase interfaces. 
The upscale transfer to moist $PV$is influenced by nonlinear waves at lowest order resulting from nonlinear buoyancy near phase interfaces. These nonlinear waves lead to coherent updrafts and downdrafts roughly aligned with fuzzy, large-scale phase boundaries identified by the
time average of a cloud indicator function. 
Statistical relationships between phase boundaries, updrafts/downdrafts and moist $PV$ are explored in flow regions dominated by moist $PV$-vortices.

\end{abstract}
% \maketitle
\bigskip
\noindent
Keywords. Atmospheric flows, condensation/evaporation
% \keywords{Atmospheric flows, condensation/evaporation}%, wave-\begin{color}{red}vortical\end{color} interactions 

% {\bf MSC Codes }  {\it(Optional)} Please enter your MSC Codes here

\section{Introduction}
\label{sec:intro}

The study of dual energy transfer (forward and inverse transfer of energy) in geophysical fluid dynamics was spearheaded by \cite{c71} in his study of quasigeostrophic flow for midlatitude dynamics.  Charney
recognized the analogy between quasigeostrophic turbulence and two-dimensional turbulence, the latter for which \cite{kraichnan67} had explained the suppression of the forward cascade of energy and the prominence of the inverse cascade of energy.  In \cite{c71}, the focus was mainly on the inhibition of the forward cascade, resulting in small-scale energy spectrum with steep scaling $k^{-3}$, where $k$ is the three-dimensional wavenumber.  
A few years later, Herring began to probe the inverse cascade regime of quasigeostrophic turbulence, both theoretically and numerically \citep{herring80,herring88}.  Herring and Metais pioneered the theoretical and numerical study of energy transfers under the fully three-dimensional Boussinesq approximation for stably stratified flows, including both geostrophic modes and waves \citep{herringmetais89,metaisherring89}.  Those studies are foundational for understanding layering in strongly, stably stratified flows observed widely in laboratory and natural settings.

During the next three decades and continuing today, a vast community of researchers has been devoted to theoretical, numerical, and laboratory studies of energy transfers, statistics, and physical space structures of the stably stratified Boussinesq equations, with and without rotation.
By varying the rotation rate and the buoyancy frequency, research spans the regimes of the rotating Boussinesq equations, including quasi-geostrophic turbulence described by the pseudo potential vorticity, strongly rotating flows, and strongly stratified flows with and without rotation \citep[e.g.,][]{m03}.

The goal of the present work is to extend the body of knowledge on Boussinesq dynamics including moisture and phase changes of water, which are fundamental for understanding
moist atmospheric convection \citep{stevens2005atmospheric} and 
moist planetary convection more generally \citep{muraetal2022}. For comparison to the results for dry Boussinesq dynamics, we re-consider the idealized set-up of nonlinear dynamics in periodic domains intended to focus on the intrinsic effects of nonlinearity.  When phase changes of water are present, a buoyancy nonlinearity accompanies the quadratic nonlinearity. Our purpose is to investigate the influence of the buoyancy nonlinearity
on inverse transfer of energy and the formation of large-scale coherent structures. 

Although both freely decaying and forced scenarios are of interest, here we focus on dynamics forced randomly at small scales, where the small-scale forcing acts as proxy for input of energy by generic unresolved processes, such as scattered convection. As background and context regarding inverse transfer in the dry Boussinesq setting, a (small) sampling of  numerical computations of inverse transfer 
%driven by small-scale forcing 
may be found in \cite{bartello95,metaisetal96,cambon01,sw02,wingate2011,marinoetal2013,brunner2014-jpo,herbert2016}.  These studies span the regimes mentioned above, namely flows that are rotation dominated or stratification dominated, and flows with strong and comparable 
rotation and buoyancy effects.  Notably, there is also an extensive literature on inverse transfer in other idealized models for dry geophysical flows such as flow on the $\beta$-plane \citep{rhines75,vallismaltrud93}, zonostrophic turbulence \citep{galperin2008,galperin2010}, and rapidly rotating Rayleigh-B\'{e}nard convection \citep{stellmachetal2014}.

In single-phase, idealized models involving slowly varying modes (zero-frequency linear eigenmodes) and fast waves (high-frequency linear eigenmodes), energy transfers may be analyzed in terms of resonances and non-resonances \citep[e.g.,][]{longuet67,lelongriley91,bartello95,sw02,leesmith07,clarkdiLeoni16}. In periodic domains, rigorous analyses for asymptotically small Rossby and/or Froude numbers are given in \cite{babinetal1997,em96,em98,me98,wingate2011}. In these limiting flows, the dry dynamics for slowly varying quantities is not influenced by the fast waves.  However, at finite parameter values, coupling between fast and slow modes via non-resonant interactions contributes to the generation of coherent structures.

Neither slow-fast eigenmode analysis nor resonant-nonresonant wave analysis is straightforward in the presence of phase changes because the buoyancy frequency changes across phase boundaries. Thus, different phases have different linear eigenmodes in Fourier space, and representation of the flow by a single Fourier linear eigenmode-basis is no longer possible.  Indeed, waves in flows with phase change are {\it nonlinear}, and evidence to date suggests that they might influence slowly varying dynamics at lowest order, even for limiting parameter values 
\citep{zhang2021fast,zhang2021effects,remondtiedrez-decomposition2024}. Using physical-space arguments, first steps to generalize fast-wave-averaging results to flow with phase changes are given in \cite{zhang2021fast}, with support from numerical computations in \cite{zhang2021effects,zhang2022convergence}.   The path forward for rigorous analysis of fast-wave-averaging with phase changes is discussed in \cite{remondtiedrez-decomposition2024}.  
We note that the presence of solid boundaries also affects slow-fast and resonant-nonresonant wave analysis.
For a channel domain, \cite{bardos2024derivation} rigorously derived a generalized quasi-geostrophic approximation, where the fast-wave correction to the slow dynamics is identified as a new resonance term. 

The PDE community has provided rigorous results for several moist systems related to the moist Boussinesq system studied in this work.  A model with water vapor and liquid water under the influence of a prescribed velocity field was analyzed in \cite{coti_zelati_temam_12,coti_zelati_fremond_temam_tribbia_13, bousquet_coti_zelati_temam_14}. 
Results for water vapor and liquid water evolving according to the primitive equations are found in \cite{coti_zelati_huang_kukavica_temam_ziane,lian_ma,temam_wu_15, teman_wang_16}.  Then followed analyses wherein liquid water is divided into cloud water (that does not fall) and rain water, that is, for `warm-rain, bulk-cloud' physics \citep{grabowski,klein_majda_06}. The case of prescribed velocity is considered in \cite{cao_hamouda_temam_tribbia,hittmeir_klein_li_titi_17}. Theorems for primitive-equation velocity are proved in \cite{tan_liu,hittmeir_klein_li_titi_20}.
Incorporation of ice into water microphysics is discussed in \cite{cao_jia_temam_tribbia}.
Recently, \cite{remondtiedrezetal2024,remondtiedrez-decomposition2024} rigorously analyzed a nonlinear 
elliptic PDE underlying slow-fast decomposition of a moist Boussinesq system  with vapor and clouds.

In an idealized Boussinesq setup, we investigate the generation of large-scale coherent structures in the presence of phase transitions, where a random force acts as a surrogate for unresolved processes and physical instabilities at small scales.  Akin to forced turbulence studies, the aim is to characterize fully nonlinear energy-transfer dynamics and the resulting quasi-statistically steady states.
The complementary literature on special solutions and moist instabilities is large, and we mention only a few studies that are closely connected to the present study, considering idealized setups of moist dynamics, including effects of nonlinear buoyancy.  A foundational study using a shallow-water model \citep{gill1982} emphasized the slower propagation speed of waves in moist regions, leading to front solutions. Extensions of the model elucidated the mechanisms for drying and precipitation fronts \citep{friersonetal2004,pauluisetal2008,bouchutetal2009}.  Moist shallow-water models have also been used to investigate the baroclinic instability of the Bickley jet \citep{Lambaertsetal2012}, 
barotropic and baroclinic instabilities of vortices \citep{rostamizeitlin2017},
and instability of hurricane-like vortices \citep{Lahaye2016,rostamizeitlin2018}. For the moist Boussinesq system considered herein, \cite{hernandezduenasetaljas2015} performed a linear stability analysis of saturated environments, with and without rainfall. In the quasi-geostrophic limit of the same Boussinesq equations, \cite{wetzel_smith_stechmann_17} analyzed the baroclinic instability and  \cite{wetzel_smith_stechmann_19_fronts} found classes of discontinuous, front-like solutions.

In Section \ref{subsec:model}, we introduce the moist Boussinesq model with asymptotically fast, warm-rain microphysics \citep{hmss13} and its energy conservation statement \citep{marsico2019energy}. Section \ref{subsec:numerical} describes the numerical simulations and discusses the slow-fast decomposition in the presence of phase changes \citep{zhang2021fast,remondtiedrez-decomposition2024}.  The results are explained in Section \ref{sec:results} by way of energy statistics; generation of coherent structures; and coupling between 
nonlinear waves and moist potential vorticity in regions characterized by frequent phase transitions.  A summary is provided in Section \ref{sec:discussion}.

% \label{sec:methods}

\section{The Moist Model}
\label{subsec:model}

\subsection{Dynamical Equations}
\label{subsec:dynamical equations}

We consider moist stably stratified Boussinesq dynamics in a frame of reference rotating about the $\hat{\bf z}$-direction, and where the buoyancy $b$ includes bulk contributions from water vapor and liquid water. Similar models have been used for many purposes 
and with varying degrees of idealization
\citep[e.g.,][]{k61,s76,b87i,cd93,sby06,ps10,hmss13,marsico2019energy,zhang2021fast,zhang2021effects,vpt19}.

The (inviscid) moist Boussinesq model is given by
\begin{subequations}
\label{eqn:original-boussinesq}
\begin{align}
&\frac{D\mathbf{u}}{Dt} + f\hat{\bf z} \times {\bf u}
= -\nabla \phi
+b \; \hat{\bf z},
\label{eqn:original-momentum}
\\
&\frac{D\theta}{Dt} + \frac{d\tilde{\theta}}{dz}w
= \frac{L_v}{c_p}(C -E),
\label{eqn:original-theta}\\
&\frac{D q_v}{Dt} + 
\frac{d\tilde{q}_v}{dz}w= - C + E,
\label{eqn:original-qv}\\
&\frac{D q_l}{Dt} - V_T\frac{\partial q_l}{\partial z}= C - E,
\label{eqn:original-ql}\\
&\nabla\cdot \mathbf{u} = 0,
\label{eqn:original-continuity}
\end{align}
\end{subequations}
where $D/Dt = \partial/\partial t + {\bf u}\cdot \nabla$ is the material derivative, and the unknown fields ${\bf u}, \phi, \theta, q_v, q_l$ are functions of space ${\bf x} = (x,y,z)$ and time $t$. Using standard notation, ${\bf u} = (u,v,w)$ is the velocity vector, $\theta$ is the potential temperature, $\phi$ is the effective pressure, $q_v$ is the mixing ratio of water vapor, and $q_l$ is the mixing ratio of liquid water.
The model buoyancy $b$ is given by 
\begin{align}
b &= g \left( \frac{\theta}{\theta_0} + R_{vd}q_v -q_l  \right),
\label{eqn:b-def-theta}
\end{align}
where $\theta_0 \approx 300$ K is the constant part of the background potential temperature, $g \approx 9.8 \mbox{ m s}^{-2}$ is the acceleration of gravity and $R_{vd} = (R_v/R_d) - 1 \approx 0.61$, and $R_d$ ($R_v)$ is the gas constant for dry air (water vapor).  
The Coriolis parameter $f$ is approximated by a constant, and consistent with mid-latitude large scales, the corresponding Rossby number is chosen $O(10^{-1})$ (see discussion at the end of this Section \ref{subsec:dynamical equations}).

%The functional form of the buoyancy $b =b(\theta, q_v, q_l)$ will be explained later in this section.

The thermodynamic variables have been decomposed into a background function of altitude $z$ and a fluctuating part, such that  $\theta^\textrm{tot} = \tilde{\theta}(z) + \theta$ and $q_v^\textrm{tot} = \tilde{q}_v(z) + q_v$. Notice that in \eqref{eqn:original-ql}, we have chosen $\tilde{q_l}(z) = 0$.  For simplicity of the numerical simulations presented herein, all non-zero background profiles are chosen to be linear in altitude $z$.

The source term $C$ represents condensation of water vapor to form liquid water, while $E$ denotes evaporation of liquid water to form water vapor.  Both $C$ and $E$ are microphysical phase-change processes that are  modeled at the macroscale \citep[e.g.,][]{grabowski}.  Since the equation for potential temperature $\theta$ is derived from conservation of energy, the source-sink term $(L_v/c_p)(C-E)$ indicates that energy is released for $(C-E) > 0$ and absorbed for $(C-E) < 0$, where the latent heat $L_v$ and specific heat $c_p$ are assumed constant, with values $L_v = 2.5 \times 10^{6} \; \textrm{J} \; \textrm{kg}^{-1}$ and $c_p = 10^3 \; \textrm{J} \; \textrm{kg}^{-1} \; \textrm{K}^{-1}.$ 
In our model, the time scales associated with condensation and evaporation are assumed asymptotically fast compared to all other time scales \citep{hmss13}, and then the difference $C-E$ is given by the vertical velocity $w$ and a prescribed saturation function of altitude $z$ (see \eqref{eqn:CminusE} and the accompanying text).  Furthermore, \eqref{eqn:original-ql} adopts a simplified representation of rainfall, with constant fall speed $V_T$. 

It is convenient to make the change of dependent variables $\theta_e^\textrm{tot} = \theta^\textrm{tot} + (L_v/c_p) q_v^\textrm{tot}$ and $q_t^\textrm{tot} = q_v^\textrm{tot} + q_l^\textrm{tot},$ because the source-sink $(C-E)$ does not explicitly appear in the equations for the equivalent potential temperature $\theta_e^\textrm{tot}$ and total water mixing ratio $q_t^\textrm{tot}$.  Then one can see that the equivalent potential temperature $\theta_e^\textrm{tot}$ is materially conserved, and the total water mixing ratio $q_t^\textrm{tot}$ is materially conserved in the absence of rain.  %Under this change of variables, the equations become
It is important to note, however, that the source-sink term and phase change information do not vanish due to the change of variables. Instead, they enter the equations through the buoyancy term \(b\), which is rewritten in terms of the new variables (see equations \eqref{eqn:buoyancy}–\eqref{eqn:bu-bs-def} and the accompanying text for further details). Under this change of variables, the equations become 
\begin{subequations}
\label{eqn:M-boussinesq}
\begin{align}
&\frac{D {\bf u}}{Dt} +f \hat{\bf z} \times {\bf u}= -\nabla \phi + b \;\hat{\bf z}, \label{eqn:u-evol} \\
&\frac{D\theta_e}{Dt} + \frac{d\tilde{\theta}_e}{dz}w = 0, 
\label{eqn:thetae-evol} \\
&\frac{D q_t}{Dt} + \frac{d\tilde{q}_t}{dz}w = V_T \frac{\partial q_l}{\partial z},
\label{eqn:qt-evol} \\
&\nabla \cdot {\bf  u} =0,
\label{eqn:incomp}
\end{align}
\label{eqn:boussinesq}
\end{subequations}
where we assume stable stratification with $d \tilde{\theta}_e/d z > 0$ and $d \tilde{q}_t/d z < 0.$
The total water mixing ratio $q_t = q_v + q_l$ combines water vapor $q_v$ and liquid water $q_l$, the latter which falls at constant speed $V_T.$  In the case $V_T=0$, all liquid water is cloud water,
while for $V_T>0$, all liquid water is rain.  In other words, the model assumes asymptotically fast autoconversion from cloud water to rain water, in addition to assuming asymptotically fast condensation and evaporation.

The Boussinesq model 
\eqref{eqn:original-boussinesq}-\eqref{eqn:b-def-theta}, or equivalently 
\eqref{eqn:b-def-theta}-\eqref{eqn:boussinesq}, partitions water as either unsaturated with $q_t^\textrm{tot} < q_{vs}^\textrm{tot}$, or saturated with $q_t^\textrm{tot} \geq q_{vs}^\textrm{tot}$, where 
$q_{vs}^\textrm{tot} = q_{vs}^\textrm{tot}(p^\textrm{tot},T^\textrm{tot})$ is the saturation mixing ratio given by the Clausius-Clapeyron relation, $p^\textrm{tot}$ is total pressure and $T^\textrm{tot}$ is total temperature. Consistent with the Boussinesq approximation, we choose 
$q_{vs}^\textrm{tot} \approx 
\tilde{q}_{vs}(z), \;q_{vs}=0$ \citep{hmss13}.  We further simplify using the background state $\tilde{q}_t = \tilde{q}_v =\tilde{q}_{vs}$ and $\tilde{q}_l = 0$, where
$\tilde{q}_t$ is a decreasing linear function of altitude. 
According to these choices, the expressions 
\begin{equation} 
q_v = \text{min}(q_t,0), \quad q_l = \text{max}(0, q_t)
\end{equation}
determine fluctuating water vapor $q_v$ and fluctuating liquid water $q_l$ from fluctuating total water $q_t$.  Then in the limit of fast water microphysics, the source term $(C-E)$ appearing in the equations for $\theta, q_v, q_l$ is given by
\begin{equation}
C-E = 
\begin{cases}
0 & \text{if}\ q_t < 0 \\
-w \; d\tilde{q}_{vs}/dz& \text{if}\ q_t \ge 0.
\end{cases}
\label{eqn:CminusE}
\end{equation}
Since $d\tilde{q}_{vs}/dz$ is a negative constant, one can see from \eqref{eqn:CminusE} that vapor is condensed from vapor to liquid in a saturated environment with $w > 0.$

For understanding and diagnosing the phase boundaries, it is helpful to write the buoyancy as
\begin{align}
    b = b_u H_u + b_s H_s, \label{eqn:buoyancy}
\end{align}
where $b_u$ and $b_s$ are the unsaturated and saturated buoyancy, respectively, and  where $H_u$ and $H_s$ are Heaviside functions
\begin{equation}
    H_u = 
    \begin{cases}
    1 & \text{if}\ q_t < 0 \\
    0 & \text{if}\ q_t \ge 0,
    \end{cases}
    \qquad
    H_s = 1-H_u.
\label{def:indicatorqt}
\end{equation}
Note that the piecewise nature of the buoyancy \eqref{eqn:buoyancy}-\eqref{def:indicatorqt} with $H_u=H_u(q_t), H_s = H_s(q_t)$ means that buoyancy is a nonlinear function.
To define $b_u$ and $b_s$, the buoyancy definition
(\ref{eqn:b-def-theta}) is rewritten in terms of $\theta_e$ and $q_t$,
which yields
\begin{equation}
    b_u = g \left( \frac{\theta_e}{\theta_0} + \left( R_{vd} - \frac{L_v}{c_p\theta_0}\right)q_t \right), \quad 
    b_s = g \left( \frac{\theta_e}{\theta_0} -q_t \right).
    \label{eqn:bu-bs-def}
\end{equation}
These types of unsaturated and saturated buoyancy variables
have also been used in other work on moist convection \citep[e.g.,][]{k61,b87i,ps10,ss17}
and are sometimes referred to as dry and moist buoyancy variables.
Notice that $b=b_u$ in unsaturated regions, and $b=b_s$ in
saturated regions, but the variables $b_u$ and $b_s$ are defined
and exist everywhere, since they are defined as functions of
$\theta_e$ and $q_t.$  
Their corresponding buoyancy frequencies are given by
\begin{equation}
    N_u^2 = g \frac{d}{dz}\left( \frac{\tilde{\theta}_e}{\theta_0} + \left( R_{vd} - \frac{L_v}{c_p\theta_0}\right)\Tilde{q}_t \right), \quad
    N_s^2 = g \frac{d}{dz}\left( \frac{\tilde{\theta}_e}{\theta_0} - \tilde{q}_t \right),
    \label{eqn:Nu-Ns-def}
\end{equation}
which may by used to define Froude numbers $Fr_u = U/(N_u H)$, $Fr_s = U/(N_sH)$ characterizing unsaturated and saturated regions of the flow, where $U$ and $H$ are, respectively, the characteristic flow-speed and height. In the triply periodic domain used here, the characteristic height and horizontal length scales are equal to each other, with $H=L=2\pi$. Along with the Rossby number $Ro = U/(f L)$, the two Froude numbers $Fr_u, Fr_s$ are necessary to characterize the possible flow regimes.  

The set-up considered here is a finite-$\epsilon$ version of the moist quasi-geostrophic regime with 
$Ro = \epsilon$, $Fr_u = O(\epsilon)$ and $Fr_s = O(\epsilon)$ 
\citep{ss17}. Our main numerical computation uses 
$Ro =\epsilon \approx 0.17$, $Fr_u \approx 0.17$ and $Fr_s \approx  0.24$ (Table \ref{table: parameter setting}). 
The dry version of this set-up with $Ro = \epsilon$, $Fr = O(\epsilon)$
would exhibit an inverse cascade of energy \citep{c71}, where $Fr = U/(NH)$, $N^2 = g \theta_0^{-1} d\tilde{\theta}/dz$, where $\theta$ is potential temperature.
%associated with the quasi potential vorticity \cite{c71}, 
Here we explore how the inverse cascade is modified by 
the presence of moisture and phase changes.

\subsection{Slow variables for strong rotation and stratification}
\label{subsec:fast-slow-variables}

The dry dynamics $(u,v,w,\theta)$ may be completely characterized by a superposition of eigenmodes obtained from the linear operator, which includes the Coriolis operator parameterized by rotation frequency $f$ (or non-dimensional $Ro$), and buoyancy operator parameterized by buoyancy frequency $N$ (or non-dimensional $Fr$) \citep{m03}.  The `slow' pseudo potential vorticity ($PV$) eigenmode does not vary in time in the limit $Fr \sim Ro = \epsilon \rightarrow 0$, and is thus associated with slowly varying nonlinear dynamics under the influence of strong rotation and stratification.  The wave eigenmodes are propagating solutions to the linear equations, with frequencies between $f$ and $N$ ($Ro^{-1}$ and $Fr^{-1}$), and are referred to as `fast' in such a decomposition.

In the moist system \eqref{eqn:boussinesq} under the influence of strong rotation and stratification (both temperature and moisture), the decomposition into fast and slow variables cannot be found from linear eigenmode analysis because the buoyancy term is {\it nonlinear} according to \eqref{eqn:buoyancy}.  Nonlinear buoyancy gives rise to nonlinear waves with frequencies that change on different sides of a phase boundary.  Nevertheless, \cite{ss17} shows how to construct a slowly varying moist pseudo potential vorticity $PV$, and a slowly varying $M$-variable which depends on water $q_t$.  Similar to the dry dynamics, the slowly varying $PV$ and $M$ variables are solutions to the nullspace of the operator arising from the rotation and (nonlinear) buoyancy terms in \eqref{eqn:boussinesq}. Furthermore, \cite{remondtiedrez-decomposition2024} shows how to construct a complete decomposition of the lowest-order nonlinear dynamics based on the $PV$-and-$M$ inversion introduced in \cite{ss17}.

Depending on whether the background environment is mostly unsaturated or mostly saturated, one may consider moist $PV$ based on either the unsaturated buoyancy frequency $N_u$ or the saturated buoyancy frequency $N_s$ \citep{ss17}.  In our simulations with significant cloud fraction, we study $PV_s$ given by  
\begin{equation}
PV_s = \hat{\bf z} \cdot (\nabla \times {\bf u})  + \frac{f}{N_s^2} \frac{\partial b_s}{\partial z},
\label{def:PV_s}
\end{equation}
where $b_s$ is the saturated buoyancy in \eqref{eqn:bu-bs-def} and $N_s$ is given in 
\eqref{eqn:Nu-Ns-def}. The slowly varying moisture variable is defined as
\begin{equation}
M = q_t + G_m \theta_e,\quad G_m = - \frac{d\tilde{q}_t}{dz} \biggl ( \frac{d\tilde{\theta}_e}{dz} \biggr)^{-1}.
\label{def:M-1}
\end{equation}

The moist pseudo potential vorticity $PV_s$ is an analog of dry pseudo potential vorticity in several ways.  First, $PV_s$ is the linear part of a nonlinear Ertel-like potential vorticity, $PV_{s,\textrm{Ertel}} = (f\hat{\bf z} + \nabla \times {\bf u}) \cdot \nabla b_s^\textrm{tot},$ for which there is a patch-integrated conservation statement \citep{kooloth2022conservation,kooloth2024}.  
Although strict material invariance is lost because of phase changes, $PV_{s,\textrm{Ertel}}$ is conserved following certain local volumes, enclosed by surfaces that are defined using two materially conserved flow quantities. Second, $PV_s$ is a slow variable because it does not change in time as $Fr_s \sim Fr_u \sim Ro \rightarrow 0$, after discarding quadratic nonlinearities \citep{zhang2021effects}.  Third, in the limit $Fr_s \sim Fr_u \sim Ro \rightarrow 0$, there is an invertibility principle starting from  $PV_s$ and $M.$  In other words, from $PV_s$ and $M$, one can recover slowly varying ${\bf u}_h, \theta_e$ and $q_t$ using geostophic and hydrostatic balance \citep{ss17,remondtiedrez-decomposition2024}.   In light of these analogous features for dry $PV$ and moist $PV_s$, one expects that $PV_s$ may be representative of large-scale, slowly varying motions in flow with phase changes between water vapor and liquid water.

As mentioned in Section~\ref{sec:intro}, asymptotic analysis of coupling between waves and slowly varying $PV_s, M$ is complicated by the presence of phase boundaries \citep{zhang2021fast,zhang2021effects,zhang2022convergence,remondtiedrez-decomposition2024}.  One goal of the simulations presented herein is to continue building intuition regarding nonlinear coupling for finite, small  Rossby and Froude numbers %on the order of $O(10^{-1})$ 
in the presence of phase changes.

\subsection{Energy conservation}
\label{subsec:energy}

For the inviscid Boussinesq equations \eqref{eqn:boussinesq}, \cite{marsico2019energy} showed that the total energy $E$ may be written in terms of $b_u, b_s,$ and chosen as
\begin{align} %\label{eq:energy-vt}
E &= KE + PE_u + PE_s + PE_M\\
&=\frac{1}{2} {\bf u} \cdot {\bf u} + \frac{1}{2} \frac{b_u^2}{N_u^2} H_u +  \frac{1}{2} \frac{b_s^2}{N_s^2} H_s + \frac{1}{2} \biggl (A_u H_u + 
\frac{1}{2} A_s H_s \biggr ){\cal{M}}^2, 
\label{eq:energy-vt}
\end{align}
where $KE$ is kinetic energy, $PE_u$ ($PE_s)$ are potential energies associated with unsaturated (saturated) domains, and $PE_M$ is a moist latent potential energy arising upon change of phase. The variable ${\cal M} = g\dfrac{N_u^2-\alpha N_s^2}{N_u^2N_s^2}M$  such that
\begin{equation}
    {\cal{M}} = \frac{b_u}{N_u^2} - \frac{b_s}{N_s^2},
\label{eqn:M}
\end{equation}
and the
coefficients $A_u, A_s$ are given by 
% \begin{equation}
% A_u = \frac{N_s^2 \alpha}{(N_u^2 - N_s^2)(\alpha N_u^{-2}+ N_s^{-2})}, \quad A_s = \frac{-N_u^2}{(N_u^2 - N_s^2)(\alpha N_u^{-2}+ N_s^{-2})},
% \end{equation}
\begin{equation}
A_u = \frac{-N_s^2 \alpha}{(N_u^2 - N_s^2)(N_s^{-2} -\alpha N_u^{-2})}, \quad A_s = \frac{-N_u^2}{(N_u^2 - N_s^2)(N_s^{-2}-\alpha N_u^{-2})},
\end{equation}
with 
\begin{equation}
\alpha = \biggl (\frac{L_v}{c_p\theta_0} - R_{vd} \biggr ).
\end{equation}
The coefficients $A_u, A_s$ are positive if $1 < (N_u/N_s)^2 < \alpha \approx 10$. 
%(in our main simulation we use 
%$(N_u/N_s)^2 = 2$ and $\alpha = 10$).
It is interesting to note that 
$PE_u, PE_s$ and $PE_M$ are all discontinuous across a phase interface, but the total potential energy 
$PE = PE_u + PE_s+ PE_M$ is continuous.  
Following from \eqref{eqn:boussinesq}, the evolution equation for $E$ is given by 
\begin{equation}
\frac{\partial E}{\partial t} + \nabla\cdot [{\bf u}(E+\phi)] 
+ \frac{1}{2}g V_T^2\frac{(\alpha +1)}{(N_u^2 - N_s^2)}\frac{\partial}{\partial z} q_l^2 = 0.
\label{energy-vt-evolution}
\end{equation}
Integrating \eqref{energy-vt-evolution} over a triply periodic domain leads to conservation of total energy
\begin{equation}
\frac{d}{dt}\int_V E \; dV = 0,
\label{volume-integrated-energy}
\end{equation}
where $V$ indicates the triply periodic domain.
For other boundary conditions, the last term in \eqref{energy-vt-evolution} is a sink of energy at the lower boundary.

The latent energy $PE_M$ (abbreviated $M$-energy) is the critical new energy term associated with change of water phase.
The $M$-energy measures the amount of energy in either or both phases available for conversion of potential energy from one phase to the other.
In this manuscript, we 
focus on the evolution of the total energy $E$ and the $M$-energy when the equations \eqref{eqn:boussinesq} are subject to a small-scale random force (input of energy).
Total energy and total energy spectra are used to assess 
%the overall inverse energy transfer and 
the accumulation of energy at large scales.
%Comparison of 
Analysis of $M$-energy and $M$-energy spectra,
%for strongly rotating flow 
with and without rain, help to diagnose how phase changes affect such inverse energy transfer. 
%the presence of phase changes affects the energy that is accumulating at large scales, and the corresponding large-scale flow structures.

\section{Methods}
\label{subsec:numerical}

\subsection{Numerical simulations}

A forced-dissipative and non-dimensional version of the 3D moist Boussinesq system  \eqref{eqn:boussinesq} 
%with two phases of water (vapor and liquid) 
is simulated in a $2\pi$-periodic domain using a dealiased, pseudo-spectral code.   We use an equivalent system formulated in terms of the unsaturated buoyancy $b_u$ and the saturated buoyancy $b_s$ instead of $\theta_e$ and $q_t$.  The systems are equivalent because $b_u$, $b_s$ are linear combinations of $\theta_e$, $q_t$ by 
\eqref{eqn:bu-bs-def}.  The formulation in terms of $b_u,$ $b_s$ allows easy connection to the terms in the energy \eqref{eq:energy-vt}, and the definition of saturated potential vorticity \eqref{def:PV_s}.
The bulk of the manuscript considers non-precipitating flow with $V_T=0.$  A limited discussion of precipitating flow with $V_T > 0$ introduces key differences between the two cases regarding energy transfer.
%and serves as prelude to future work.  
The differences are highlighted using a like-for-like comparison in the periodic domain, while recognizing the drawbacks of this setup for precipitating dynamics.

To focus on the inverse transfer of energy to large scales, a random forcing is applied at small scales (high wavenumbers), and a hyperviscosity limits dissipative effects to scales smaller than the forcing scales.  The focus on large-scale dynamics allows for modest resolutions.
Calculations with spatial resolutions $192 \times 192 \times 192$ are sufficient for our purposes because we have verified that late-time energy ratios change only by a few percent or less when increasing the resolution to $256 \times 256 \times 256 $ (see Tables \ref{table3}-\ref{table4}).
After transferring the physical space equations into Fourier space, a third-order  Runge-Kutta time-stepping scheme solves the coupled system of ODEs resulting from discretization of the wavevector.  
%Linear 
The rotation and buoyancy terms are treated explicitly, and the nonlinear advection terms are calculated in physical space with the discrete Fourier transform algorithms of the FFTW software package (http://www.fftw.org/). A pressure-solver enforces the incompressibility constraint, and linear dissipation/diffusion terms are included using an integrating factor.

\subsection{
Energy spectra}

In order to examine the distribution of energy at different scales, we compute shell-integrated energy spectra $E(k,t)$ where $k = ({\bf k}\cdot {\bf k})^{1/2}$.  In the discrete case  $E(k_i,t)$ is calculated by summing the energy of all modes with wavenumbers in the shell $(i - 1) \Delta k  < k_i \leq  i  \Delta k$,  $\Delta k = (2 \pi)/L$, where $L=2\pi$ is length of the box and $i = 1 \cdots k_m$ (for spatial resolution is $192^3$, the maximum wavenumber $k_m = 64$ because of the $2/3$-rule for dealiasing).  Spectra may be computed for the total energy, and well as certain parts of the total energy, such as the $M$-energy.  We are particularly interested in the latter to diagnose how potential energy exchange between phases affects energy transfer to large scales.

\subsection{Hyperviscosity}

A hyperviscosity/hyperdiffusivity is used in \eqref{eqn:u-evol}-\eqref{eqn:qt-evol} to induce dissipation/diffusion only at the smallest scales. For example, in the momentum equation, the hyperviscosity acts on the velocity ${\bf u}$ according to the expression
\begin{equation}
\label{eqn:hyperform}
(-1)^{p+1} \nu(\nabla^2)^p {\bf u},
\end{equation}
where we use $p = 8$. The coefficient $\nu$ has the structure
\begin{equation}
\label{eqn:hypercoeff}
    \nu = 2.5 \biggl (\dfrac{E(k_m,t)}{k_m} \biggr )^{1/2}k_m^{2-2p},
\end{equation}
where $k_m$ is the highest available wavenumber and $E(k_m,t)$ is the kinetic energy in 
the wavenumber shell associated with $k_m$.

\subsection{Forcing}

To make the small scale forcing term a generic function,
the Fourier coefficients are treated as independent
Gaussian random variables with mean zero. The variance
for each wavenumber is assigned according to the
spectral density function
%For all cases in Section \ref{sec:result}, we consider decay from balanced large-scale, random initial conditions. 
%For all variables 
%spectral density for all %variables $(u,v,w,\theta_e,q_t)$ is a Gaussian function given by 
\begin{equation}
F(k) = \epsilon_f  \dfrac{\exp (-0.5(k- k_f)^2/s^2)}{(2 \pi)^{1/2} s},
\qquad\mbox{for}\quad k\in [k_f - a,k_f + a],
\label{eqn:gaussianspectral}
\end{equation}
where $s=1$ characterizes the spread in $k$-space, $k_f$ is the peak forcing wavenumber, and $\epsilon_f$ is an $O(1)$ coefficient.
Typically, the spectrum $F(k)$ is truncated to include only a small number of wavenumbers, for example, for resolution $192^3$ Fourier modes and $k_f = 36$, we use $a=6$ to truncate the force for $k < 30$ and $k > 42$. Such a random function is used to force the linear eigenmodes of the purely unsaturated version of equations~(\ref{eqn:M-boussinesq}) \citep{zhang2021fast}, thereby inputting energy into the variables $(u, v, w, b_u)$, with $b_u$ defined by (\ref{eqn:bu-bs-def}). Forcing $b_u$ across the entire domain implies injecting statistically equal amounts of small-scale random water vapor and liquid water. We then study the energy transfer to wavenumbers $k< 30$, as well as the structure of diagnostic quantities involving 
$(u, v,w, b_u, b_s)$ (equivalently 
$(u, v, w, \theta_e, q_t)$).

%Consequently, each variable $(u, v, w, \theta_e, q_t)$ is provided a small-scale source term in the wavenumber range $30 < k < 36,$ and one may study energy transfer to wavenumbers $k< 30$, as well as the structure of diagnostic quantities involving $(u, v, w, \theta_e, q_t)$.

\subsection{Non-dimensional parameters}

Based on the energy input rate $\epsilon_f$ and the peak wavenumber $k_f$ of the force, there is a Rossby number and two Froude number defined by
\begin{equation}
Ro = \dfrac{(\epsilon_f k_f^2)^{1/3}}{f}\quad 
    Fr_u = \dfrac{(\epsilon_f k_f^2)^{1/3}}{N_u}, \quad Fr_s = \dfrac{(\epsilon_f k_f^2)^{1/3}}{N_s}.
\end{equation}
In our simplified Boussinesq system, the frequencies $N_u$ and $N_s$ (defined by \eqref{eqn:Nu-Ns-def}) are constants because we have assumed constant background gradients $d \tilde{\theta_e}/dz$ and $d \tilde{q_t}/dz$ in equations (\ref{eqn:thetae-evol}) and (\ref{eqn:qt-evol}).
To provide physical context, in the midlatitude atmosphere on synoptic scales, the lapse rate is  approximately $d\tilde{\theta}/dz\approx 3$ K km$^{-1}$ and the decrease in water vapor with altitude is approximatley 
$d\tilde{q}_v/dz\approx -0.6$ g kg$^{-1}$ km$^{-1}$. These numbers imply that the relation $N_u^2 \approx 2N_s^2$ remains within a physically reasonable regime, consistent with a saturated buoyancy frequency $N_s$ that is lower than the unsaturated buoyancy frequency $N_u$. For simplicity, in our simulations, we use $(N_u/N_s)^2 = 2$ to set the Froude numbers $Fr_s = \sqrt{2} Fr_u$. One may also introduce Burger numbers for the  unsaturated and saturated regions, respectively, defined as 
\begin{equation}
    B_u = \dfrac{N_u^2}{f^2}, \quad B_s = \dfrac{N_s^2}{f^2}.
\end{equation}
Table \ref{table: parameter setting} shows the parameter settings for the geophysical flow regime considered in the simulation that is central in this work.

\begin{table}
    \centering
    \begin{tabular}{cccccccc}
        \hline
        $N_u$ & $N_s$ & $f$ & $B_u$ & $B_s$ & $Fr_u$ & $Fr_s$ & $Ro$  \\
        \hline
        &  &  &  &  &  &   \\
        50 & 35.36 & 50 & 1 & 0.5  &0.1664 &0.2353 & 0.1664\\
        &  &  &  &  &  & \\
    \end{tabular}
\caption{Frequencies and non-dimensional parameters for the main simulation with resolution $N=192$ and forcing spectrum \eqref{eqn:gaussianspectral} using $s=1$, $k_f = 36$, $\epsilon_f=O(1)$, $a=6$.}
\label{table: parameter setting}
\end{table}

Lastly, the rainfall term in 
\eqref{eqn:qt-evol} with constant speed $V_T \geq 0$ incorporates precipitation effects in a simple and straightforward manner. Using a reference vertical velocity $W$, the non-dimensional rainfall speed $V_r = V_T/W$ will be chosen either zero or unity. 
The value $V_r=0$ corresponds to a non-precipitating flow; the value $V_r=1$ is appropriate for 
$W \approx V_T$.

\subsection{Time scales}

Time steps in the numerical simulations are chosen small enough to satisfy the CFL condition, and to simultaneously resolve the fast-wave oscillations.  For the CFL condition, the time step $\Delta t$ satisfies $\Delta t = {\rm CFL}/(|{\bf u}|_m k_m)$, where $k_m$ is the highest available wavenumber, $|{\bf u}|_m$ is the  maximum  magnitude  of  the  velocity  field and CFL = 0.9.
To resolve the waves, each half-period is sampled at least 5 times, according to the condition $\Delta t = \pi/(5 \sigma_{max})$, where $\sigma_{max}$ is the maximum frequency of the waves. In non-dimensional form, the wave frequencies separately associated with unsaturated and saturated regions are given by~\citep{zhang2021effects}
\begin{equation}
\sigma_u({\bf k}) = \dfrac{(Fr_u^{-2}k_h^2 + Ro^{-2}k_z^2)^{1/2}}{k} ,\hspace{1cm} \sigma_s({\bf k}) = \dfrac{(Fr_s^{-2}k_h^2 + Ro^{-2}k_z^2)^{1/2}}{k}.
\label{eqn:frequency_non_dim}
\end{equation}
Therefore, with parameter choices given in Table \ref{table: parameter setting}, the maximum wave frequency is 
$\sigma_{max}= \max{(\sigma_u, \sigma_s)}$ $ = \max{(Ro^{-1}, Fr_u^{-1}, Fr_s^{-1})} = Ro^{-1}$. Upon computing $\Delta t$ based on the CFL and wave conditions, we choose the smaller of the two time steps. Finally, 
%the total time $t$ in the %model equations 
time will be non-dimensionalized by $T = (\epsilon_f k_f^2)^{-1/3}$, such that reported times $t = T^{-1}\sum_{n}  \Delta t$,
where $n$ is the number of time steps.

\subsection{Cloud fraction and analysis of phase boundaries}
\label{subsubsection:time-averages}

The cloud indicator $H_s({\bf x},t) = H_s(q_t)$ is a dynamically evolving quantity influenced by the Rossby, Froude and Burger numbers as well as by the presence or absence of rain.
The indicator $H_s$ has the value unity if the position ${\bf x}$ is saturated such that $q_t$ is given by $q_t = q_l$.  Otherwise $H_s$ is zero at locations where the flow is unsaturated with $q_t=q_v$. Locations where $H_s=1$ are associated with clouds $(V_T=0)$ or rain $(V_T> 0)$, and hence the name `cloud indicator.'
The `cloud fraction' is the $L^1$ norm of the cloud indicator $H_s(q_t)$, that is, the number of points with $H_s=1$ divided by the total number of points. For the runs with $N=192$ and $Fr_s/Ro = \sqrt{2}$, the cloud fraction reaches an approximate steady-state value of $59\%$ for both the no-rainfall ($V_r=0$) and rainfall cases $(V_r=1$).

In subsequent analysis based on time averages, the cloud indicator will be used to infer information about the frequency of phase changes at a particular location ${\bf x} = (x_0,y_0,z_0)$. 
We introduce a time average of any flow quantity $f({\bf x},t)$ over a window $[t_1,t_2]$ with duration approximately 10\% of the total simulation time.  Using brackets to denote the time average, we define
\begin{equation}
\langle f \rangle ({\bf x}) = %\lim\limits_{t \rightarrow \infty} 
\frac{1}{\tau} \int_{t_1}^{t_2} f({\bf x},s)ds,
\label{eqn:bracketavg}
\end{equation}
where $\tau = t_2-t_1$, with $\tau$ chosen towards the end of the simulation when energy has accumulated at large scales ($t_1 \approx 145$ and $t_2 \approx 160$). The interpretation of $\langle H_s \rangle$ can be understood as follows: if the value $\langle H_s(x_0,y_0,z_0) \rangle$ is close to zero or unity, then the position ($x_0,y_0,z_0$) is distant from a phase interface during the time window. Values of $\langle H_s(x_0,y_0,z_0) \rangle \approx 0.5$ suggest that the position ($x_0,y_0,z_0$) undergoes frequent phase transitions, characteristic of a position near a phase boundary.

\begin{comment}
\begin{figure}
\centering
\includegraphics[width=0.55\textwidth]{../fig/revision_june6/wpdf_3d_norainfall_timeave.png} 
\caption{{\color{blue}Probability density function (pdf) and summary statistics of $\langle w \rangle$, computed over the same time-averaging window as in Figure~\ref{fig:q_t_3d}, for the no-rainfall case.}}
\label{fig:pdfs of w}
\end{figure}
\end{comment}

In a broader context, the time-averaging operator is applied to other key physical quantities such as total water content $q_t$ and vertical velocity $w$, among others. Application of the averaging \eqref{eqn:bracketavg} effectively dampens high-frequency oscillations and fluctuations, thereby emphasizing low-frequency, slowly evolving physical structures. Consequently, analysis of time-averaged quantities using \eqref{eqn:bracketavg} uncovers key features associated with large-scale coherent structures.

\section{Results}
\label{sec:results}

Here we present the principal findings found from our numerical experiments, wherein rotation and stratification effects are strong and of comparable significance.
Results presented in Sections 
\ref{sub:energy}-\ref{subsec:dynamicwaves} pertain to the case $N_u = \sqrt{2} N_s$, $B_u=1$, $B_s=0.5$, including comparison to dry runs
(see Table \ref{table3}).  
Section \ref{sub-ODE} explains the nonlinear nature of the waves using an ODE model. Robustness studies are presented in Section \ref{subsec:NuNsimpact}.

\begin{figure}
\centering
\includegraphics[width=0.49\textwidth]{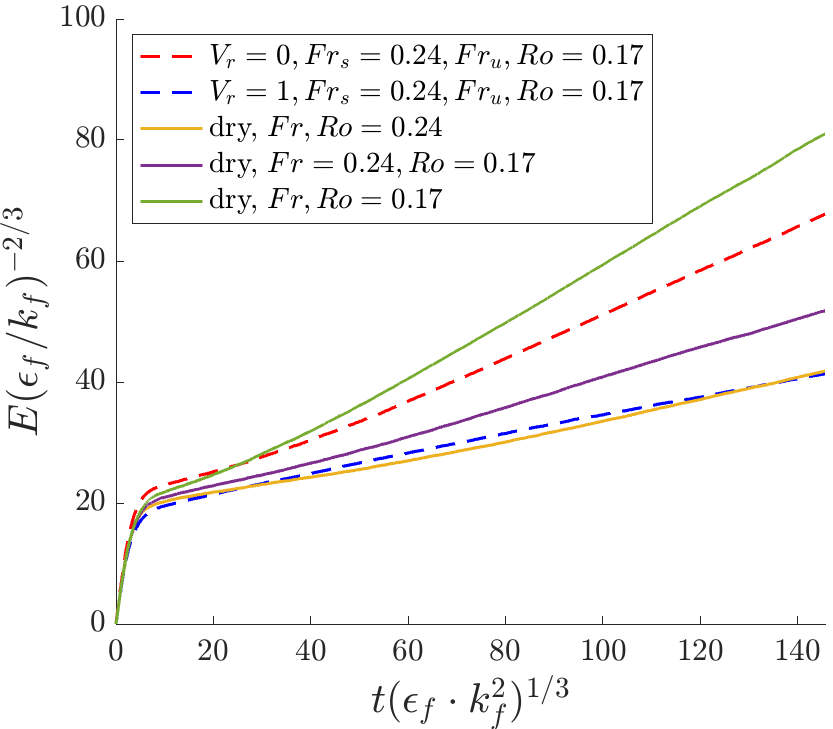}
\includegraphics[width=0.49\textwidth]{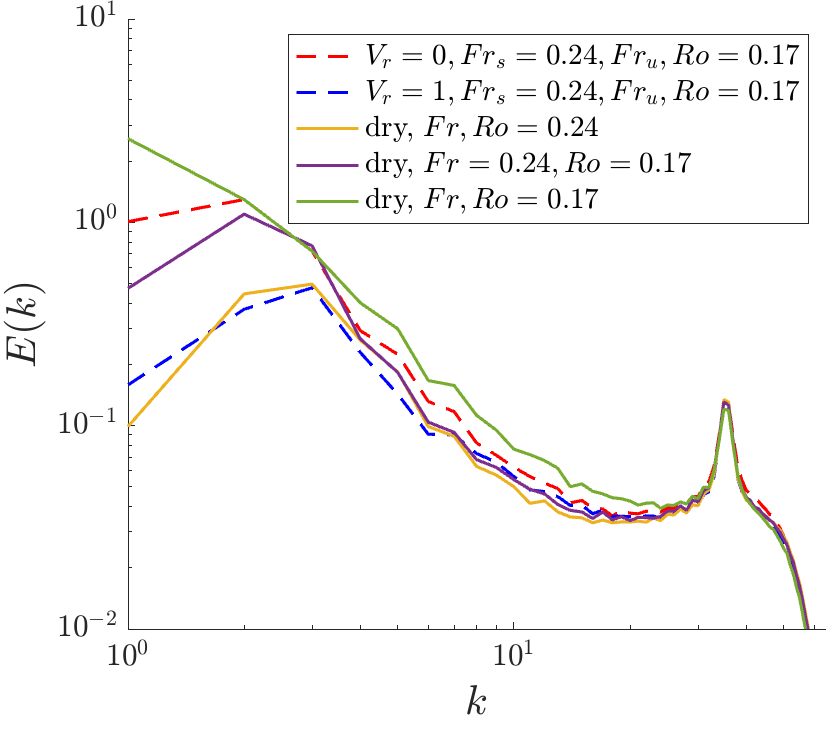}

\caption{Left: Evolution of total energy from $t=0$ to $t = 145$ in runs with resolution $N=192$. Right: Energy spectra at the end of the simulation $E(k, t = 145)$. The dashed red and blue curves indicate moist flow with, respectively, $V_r=0$ and $V_r=1$.  The solid curves indicate dry dynamics (see the legend).
}
\label{fig:energy_1}
\end{figure}

\subsection{Energy and energy spectra}
\label{sub:energy}

The evolution of the total energies 
$E(t)$ are shown in Figure \ref{fig:energy_1} (left) to assess the rate of energy transfer to scales larger than the forcing scale.  One can see a sizable difference in the growth rates for the non-precipitating case (dashed red curve) and the precipitating case (dashed blue curve).  At time $t=145$ in the run with resolution $N=192$, there is roughly 1.7 times more energy in the non-precipitating flow than in the precipitating flow (see also Table \ref{table3}). Examination of spectra in Figure \ref{fig:energy_1} (right) shows that the energy difference is mainly at small wavenumbers (large scales) with wavenumber $k \leq 10$ (compare the dashed red spectrum for non-precipitating flow to the dashed blue spectrum for precipitating flow).

Tables \ref{table3} and \ref{table4} for resolutions $N=192$ and $N=256$, respectively, provide additional quantitative information regarding the total, kinetic and potential energies in the models with phase changes, as well as in the dry model (Table \ref{table3} only).
For simplicity of notation, we use the same symbol $E(t)$ to refer to the dry energy \citep{m03} and the moist energy given by \eqref{eq:energy-vt}.  
Similarly $PE$ refers to either dry potential energy, or moist potential energy. 
%In the dry model, $PE$ is associated with the dry buoyancy.
In the moist model, recall that $PE$ has contributions from unsaturated and saturated buoyancies, as well as a latent-energy component from phase changes
(see \eqref{eq:energy-vt}).

Table \ref{table3} includes three dry runs with (i) $Ro = 0.17, Fr = 0.17, Bu = 1$; (ii) $Ro = 0.17, Fr = 0.24, Bu = 0.5$; and (iii) $Ro=0.24, Fr = 0.24, Bu = 1.$  The first dry run with $Bu=1$ has the smallest $Ro=Fr=0.17$, is closest to the quasi-geostrophic limit $Fr \sim Ro \rightarrow 0$, and thus accumulates the most energy at large scales via inverse transfer by vortical modes \citep[see Figure \ref{fig:energy_1} and][]{bartello95,herbert2016}. 
The third dry run with $Bu=1$ has the largest $Ro=Fr=0.24$, is farthest from the quasi-geostrophic limit $Fr \sim Ro \rightarrow 0$, and thus accumulates the least energy at large scales.  Case (ii) with $Ro = 0.17, Fr = 0.24, Bu = 0.5$ is in between, as can be seen in Figure \ref{fig:energy_1}.  In all the dry runs, the energy ratios are roughly $70\%$ kinetic energy and $30\%$ potential energy at the latest time $t = 145.$  

For the moist cases presented in Figure \ref{fig:energy_1} and the last two rows of Table \ref{table3}, we see that the no-rainfall case has total energy level in between dry cases (i) and (ii).  This is consistent with the fact that the unsaturated (vapor) phase has $Ro = 0.17, Fr = 0.17, Bu = 1$ as in dry case (i), while the saturated (liquid) phase has $Ro = 0.17, Fr = 0.24, Bu = 0.5$ as in dry case (ii).  Thus the impeding effect of phase changes on inverse energy transfer can partly be attributed to a higher Froude number in saturated regions of the simulation with phase changes.
By allowing rainfall in the moist dynamics, the kinetic energy $KE$ drops by 5.2\%. Evidently, rainfall has an additional impeding effect on inverse transfer of energy to large scales, as will be revisited in the next section.

\begin{table}

    \centering
    \begin{tabular}{ccccccccc}
        \hline
        $B\ |\ B_u, B_s$  & $Fr\ |\ Fr_u, Fr_s$ & $Ro$ & Model & $E$ & $KE/E$ & $PE/E$ & $PE_M/PE$ & $PE_M/E$\\
        \hline
        $1$ & $0.1664$ & 0.1664 & Dry & 87.9479 & 72.40\% & 27.60\% & N/A  & N/A \\
        $0.5$ & $0.2353$ & 0.1664 & Dry & 56.5842 & 72.82\% & 27.18\% & N/A  & N/A \\
        $1$ & $0.2353$ & 0.2353 & Dry & 41.8000 & 71.25\% & 28.75\% & N/A  & N/A \\
      $1, 0.5$   & $0.1664, 0.2353$ & 0.1664 &  No rainfall & 67.7707 & 68.20\% & 31.80\% & 19.55\% & 6.22\%\\
      $1, 0.5$  & $0.1664, 0.2353$ & 0.1664 &  Rainfall & 41.3300 & 63.05\% & 36.95\% & 39.04\% & 14.42\%\\ \hline
    \end{tabular}
    \caption{Energy and energy ratios at \( t = 145 \) using resolution \( N = 192 \). Note that in the dry model, we only have a single Burger number \( B \) defined as \( B = \frac{N^2}{f^2} \), where \( N \) is the dry buoyancy frequency, and a single Froude number \( Fr \) defined as \( Fr = \frac{(\epsilon_f k_f^2)^{1/3}}{N} \).
}
\label{table3}

\end{table}

\begin{table}
    \centering
    \begin{tabular}{ccccccccc}
        \hline
        Run & Model & $E$ & $KE/E$ & $PE/E$ & $PE_M/PE$ & $PE_M/E$\\
        \hline
         $B_u = 1, B_s = 0.5$  & No rainfall & 61.5175 & 69.89\% & 30.11\% & 22.65\% & 6.82\%\\
         $B_u = 1, B_s = 0.5$ & Rainfall & 35.2297 & 63.26\% & 36.74\% & 39.97\% & 14.68\%\\ \hline
    \end{tabular}
    \caption{Energy and energy ratios at $t
    %t(\epsilon_f\cdot k_f^2)^{\frac{1}{3}}
    = 205$ using resolution $N=256$.} 
    \label{table4}
\end{table}

\subsection{The role of the moist latent potential energy}
\label{subsec:M-energy}

Here we scrutinize the moist latent potential energy (the $M$-energy) that arises 
%only 
in the presence of phase changes and rainfall.
It is noteworthy that the $M$-energy does not directly impact energy transfers within each phase \citep{marsico2019energy}. Its main role is to exchange energy between $PE_u$ and $PE_s$, and this exchange occurs exclusively at phase interfaces. 

Figure \ref{fig:M-energy} {(left) illustrates the time tendency of the ratio of $M$-energy to total energy.
The solid curve (without rainfall) shows an initial contribution of roughly 27\%, decreasing to less than 10\% beyond a time $t \approx 80.$ The dashed curve (with rainfall) reaches a plateau of 15\% during a short start-up time of $0 < t < 8.$ The late-time $M$-spectra in Figure \ref{fig:M-energy} (right) reinforce the more significant contribution from 
%large-scale 
$M$-energy for $t > 80$, as energy is transferred to large scales, when liquid water is allowed to fall as (dashed).  Rainfall enhances latent $M$-energy at large scales, thereby reducing the inverse transfer of kinetic energy, and diminishing the amount of total energy that is accumulating at large scales.
%Consequences on physical-space structure are explored in 
%Section (\ref{subsub:physical-structure-N/F-1}).

\begin{figure}
    \centering
\includegraphics[width=0.45\textwidth]{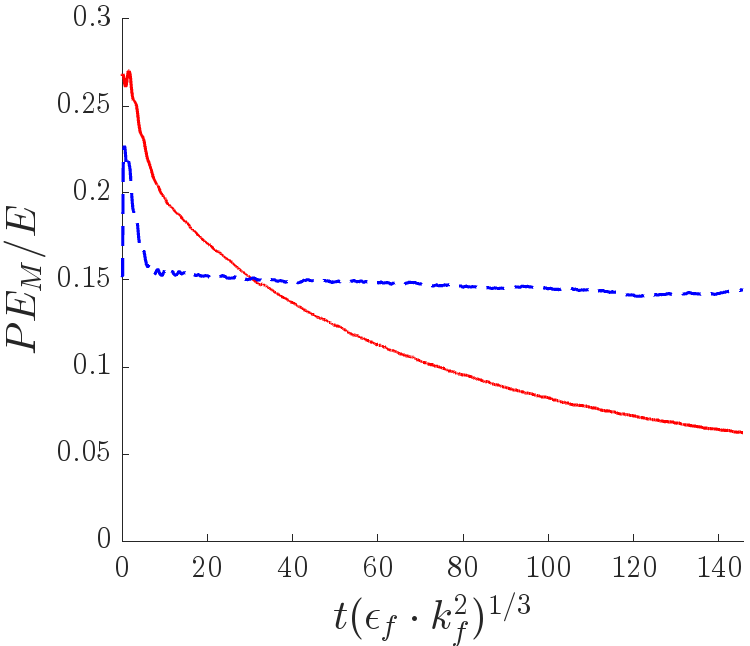}
\includegraphics[width=0.45\textwidth]{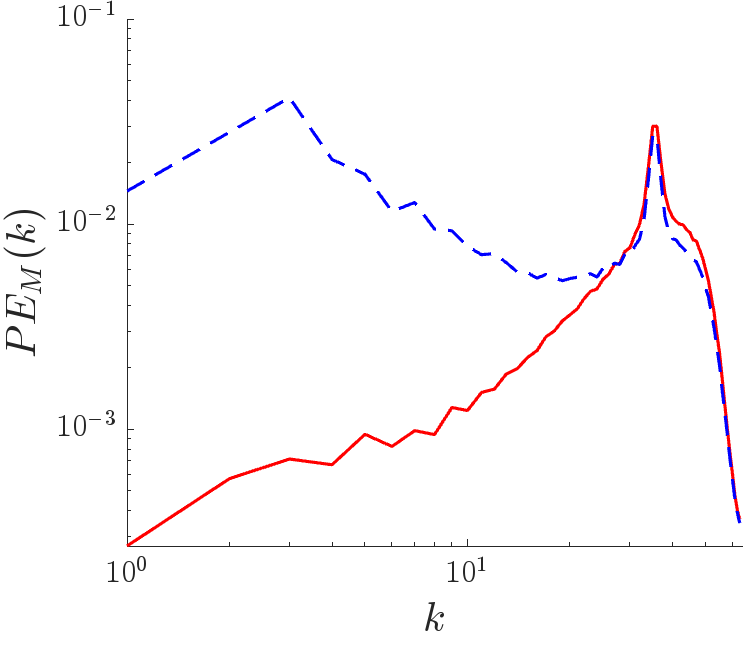}
\caption{Left: Evolution of the ratio of M-energy to total energy from $t=0$ to $t = 145$ in the run with resolution $N=192$. Right: Energy spectra at the end of the simulation $PE_M(k, t = 145)$. The solid curves indicate non-precipitating flow ($V_r=0$) and the dashed curves indicate  precipitating flow ($V_r=1$).}
\label{fig:M-energy}
\end{figure}

Above we have demonstrated the statistical significance of phase boundaries for the development of large scale flows. In the next two subsections, we analyze their significance for the  large-scale physical-space structures that arise from self-organization of randomly forced fluctuations, for both the no-rain and rainfall scenarios.  A main objective is to investigate connections between regions of high-frequency phase transition (indicated by $\langle H_s \rangle \approx 0.5$ as discussed in Section~\ref{subsubsection:time-averages}); the 
slowly varying component of vertical velocity $\langle w \rangle$ (coherent updrafts and downdrafts); and the saturated potential vorticity $PV_s$ (which is, itself, slowly varying). 

\begin{figure}
\centering

\includegraphics[width=0.45\textwidth]{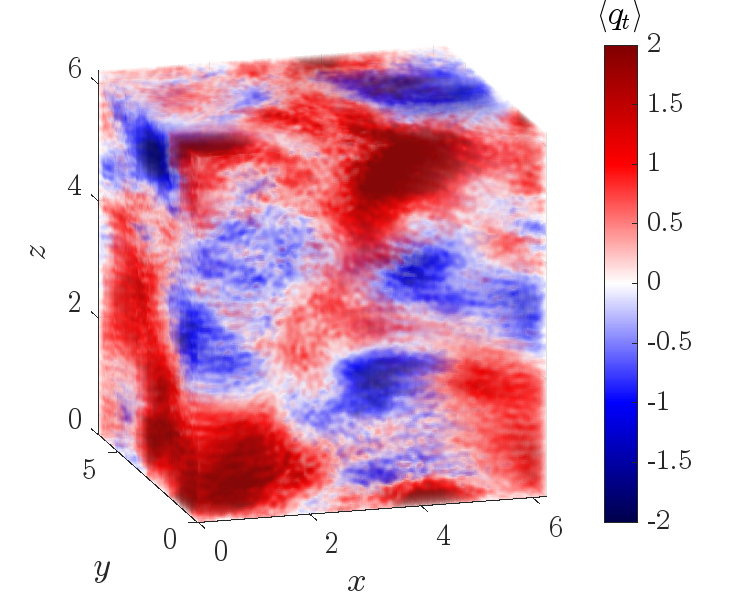}
\includegraphics[width=0.45\textwidth]{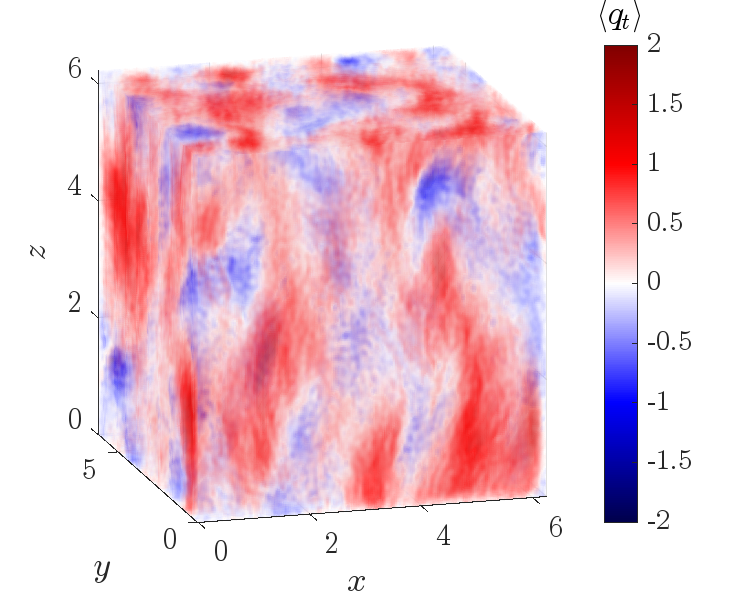}

\includegraphics[width=0.45\textwidth]{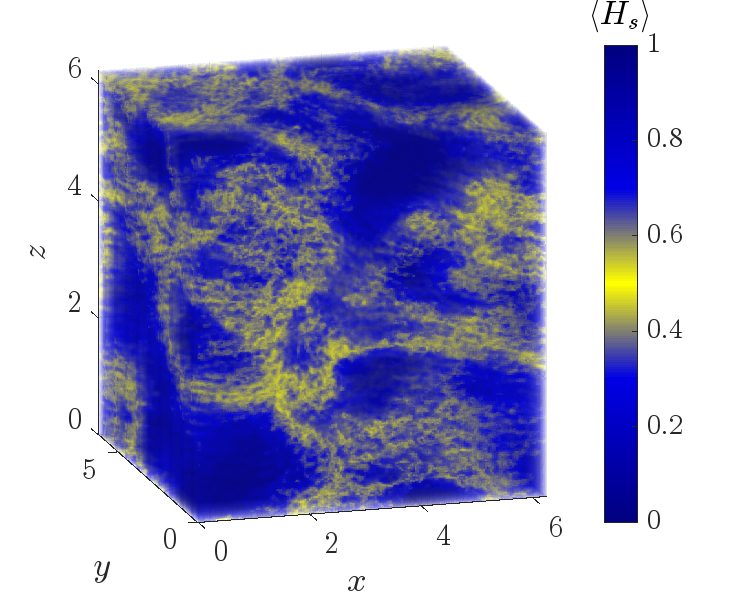}
\includegraphics[width=0.45\textwidth]{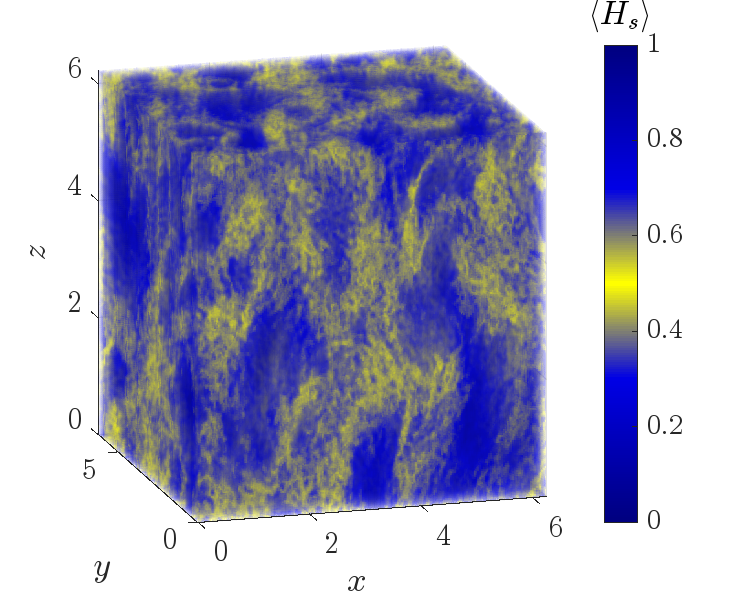}
   
\caption{Time-averages of total water $\langle q_t\rangle$ (top row) and the Heaviside function $\langle H_s \rangle$ (bottom row). The first column shows non-precipitating flow with $V_r=0$ and the second column shows precipitating flow with $V_r = 1$. Time average are performed towards the end of each simulation with resolution $N=192$, in the approximate interval $145 \leq t \le 160$. Yellow fuzzy phase boundaries in the second row are associated with areas of frequent change of water phase and values $\langle H_s \rangle \approx 0.5.$
}\label{fig:q_t_3d}
\end{figure}

%\subsection{Time-averaged, 3D physical-space flow structures}
\subsection{Time-averaged, 3D physical space water and cloud indicator}
\label{subsub:physical-structure-N/F-1}

The purpose of this subsection is to provide qualitative and quantitative intuition about the late-time water $\langle q_t\rangle$ and the phase boundaries $\langle H_s \rangle$ in the 3D domain. To do so, we examine 3D visualizations along with probability density functions (pdfs).  All 3D renderings and pdfs are time averages over the (approximate) window $145 \leq t \leq 160$ (see Figure \ref{fig:energy_1}). 

\begin{figure}
    \centering
    \includegraphics[width=0.45\textwidth]{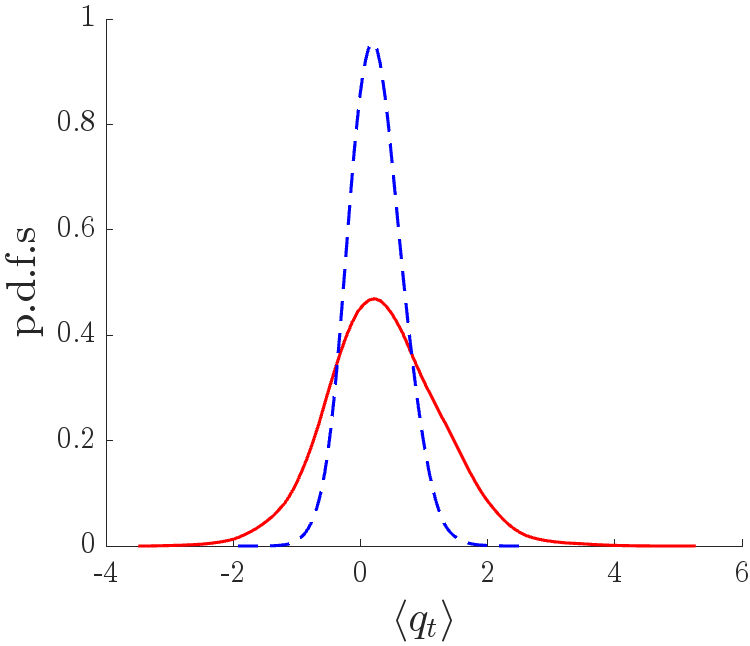}
    \includegraphics[width=0.45\textwidth]{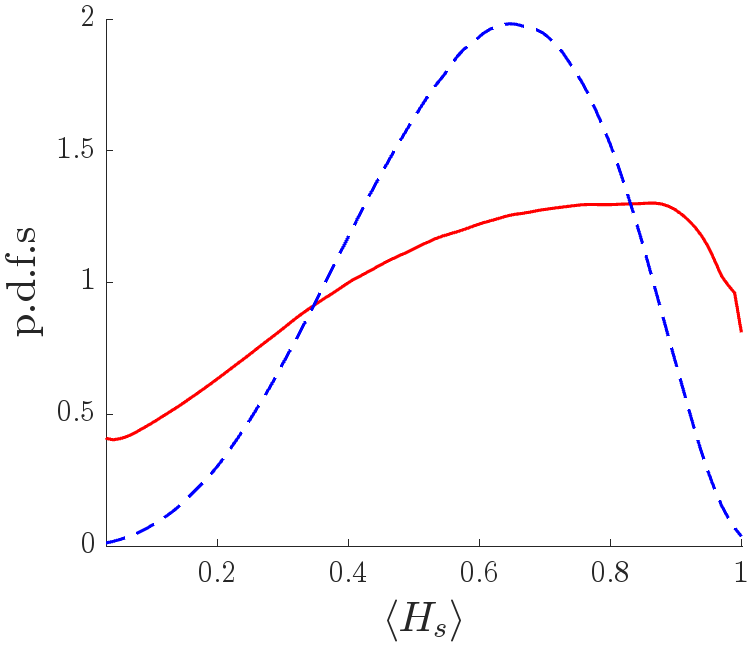}
\caption{Probability density functions (pdfs) of $\langle q_t \rangle$ (left) and 
$\langle H_s \rangle$ (right) for the same time averaging window as in Figure \ref{fig:q_t_3d}. Solid curves are without rain and dashed curves are with rain. 
Rainfall induces a shift in the pdf of $\langle q_t \rangle$ towards the saturation threshold $q_{vs} = 0$, indicating fewer points in the domain that remain purely unsaturated or purely saturated during the averaging time window.  Consequently, the cloud indicator $\langle H_s \rangle $ has fewer points associated with values zero and unity.
} 
\label{fig:pdfs}
\end{figure}

Figure \ref{fig:q_t_3d} (first row) shows the 3D physical structure of $\langle q_t \rangle$, where red regions with $\langle q_t \rangle \geq 0$ indicate clouds containing liquid water ($V_r=0$) or areas with rain ($V_r=1$), while blue regions with $\langle q_t \rangle< 0$ contain only water vapor. (Recall that $q_t$ is a fluctuation from the large background profile $\tilde{q}_t(z) = \tilde{q}_{vs}(z)$.) For the non-precipitating case with $V_r=0$ (top left), the overall large-scale structure is not dominated by vertical columns or horizontal layers, but rather exhibits both vertical and horizontal coherence, as one expects in the parameter regime where rotation and stratification are comparable. Sizable regions of dark red and dark blue indicate, respectively, relatively high liquid water content (dark red), and low vapor content (dark blue). When liquid falls as rain with speed $V_r = 1$ (top right), then the structure of the water appears more columnar and the amplitude of water fluctuations is decreased, with no dark red or dark blue areas visible.
The `fuzzy' (or smeared out) phase boundaries associated with change of sign in $\langle q_t \rangle$ are apparent in the rendering of $\langle H_s \rangle$, as shown by the yellow regions in the second row of Figure \ref{fig:q_t_3d}.  

According to Figure \ref{fig:M-energy}, there is more $M$-energy in the case when liquid falls as rain (bottom right), compared to the no-rainfall case. To relate Figure \ref{fig:M-energy} to Figure \ref{fig:q_t_3d}, we plot the pdfs of $\langle q_t \rangle$ and $\langle H_s \rangle$ in Figure \ref{fig:pdfs}.  It becomes apparent that rainfall induces a shift in the distribution of $\langle q_t \rangle$ towards the saturation threshold $q_{vs} = 0$ (left dashed), which means that there are fewer points in the domain that remain purely unsaturated or purely saturated during the averaging time window.  Consequently, the cloud indicator $\langle H_s \rangle $ for the rainfall case (right dashed) has fewer points associated with values zero and unity, and more points with values $\langle H_s \rangle \approx 0.5$, indicating frequent change of phase over more of the domain.  

In the remainder of the manuscript,
%following Sections \ref{sub-column}-\ref{subsec:NuNsimpact}, 
we focus on the no-rainfall case $V_r=0$ because it is the simplest case that illustrates the effects of the piecewise buoyancy, and because it is straightforward to interpret in the context of triply periodic boundary conditions. Interpretability of precipitating dynamics is complicated by the periodic boundary condition, in which the rain exits at the bottom and re-enters at the top. 
For completeness, Appendix \ref{subsec:nonprecipvsprecip} presents an analog of Figure \ref{fig:heightave-pv-h-w}, but with $V_r=1$ instead of $V_r=0$. 
%Further analysis of precipitating flows will be pursued elsewhere using more realistic setups, for example,  incorporating boundary effects.

\begin{comment}
{\color{blue} To better focus the study on the intrinsic effects of nonlinear buoyancy on inverse energy transfer, we choose to eliminate the influence of boundary conditions. Although periodic boundary conditions have inherent limitations and cannot fully represent real-world scenarios, we briefly revisit the rainfall case $V_r=1$ in Appendix \ref{subsec:nonprecipvsprecip} for completeness, leaving a more in-depth study of rainfall to future work.}
\end{comment}

% In the following Sections \ref{sub-column}-\ref{subsec:NuNsimpact}, we focus on the no-rainfall case $V_r=0$ because it is the simplest case that illustrates the effects of the piecewise buoyancy, and because it is straightforward to interpret in the context of triply periodic boundary conditions. For completeness, we revisit the rainfall case $V_r=1$ briefly in  Appendix \ref{subsec:nonprecipvsprecip}, leaving a more in-depth study of rainfall to future work.

\begin{figure}
    \centering
    \includegraphics[width=0.49\textwidth]{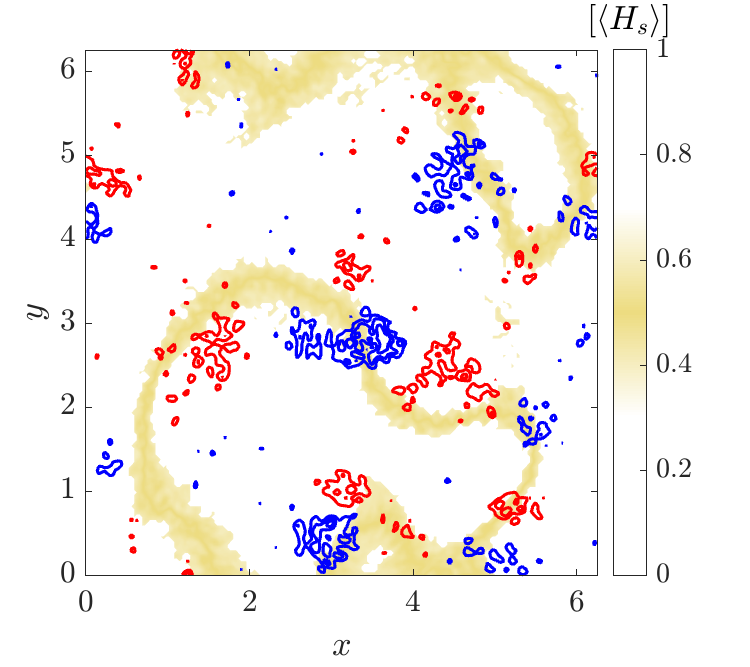}
\includegraphics[width=0.49\textwidth]{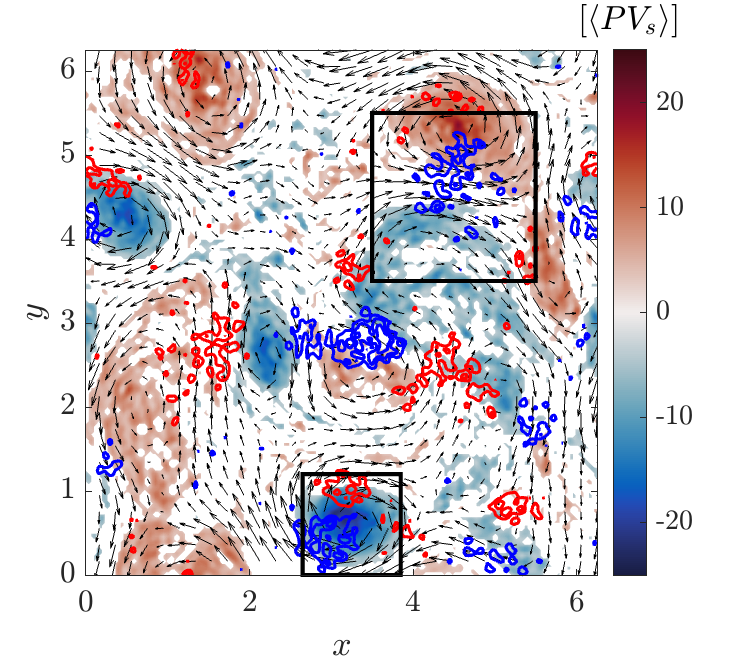}
\includegraphics[width=0.49\textwidth]{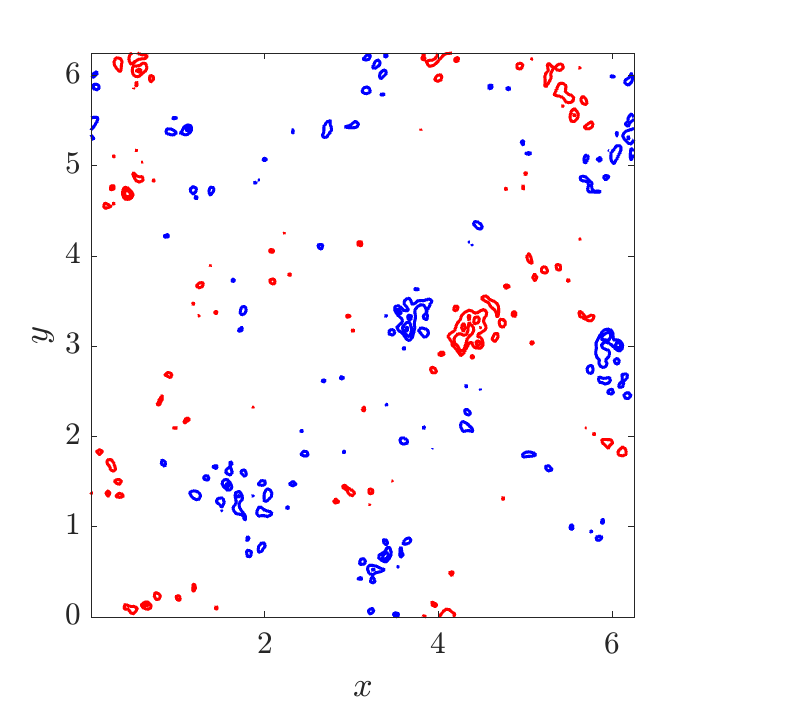}
\includegraphics[width=0.49\textwidth]{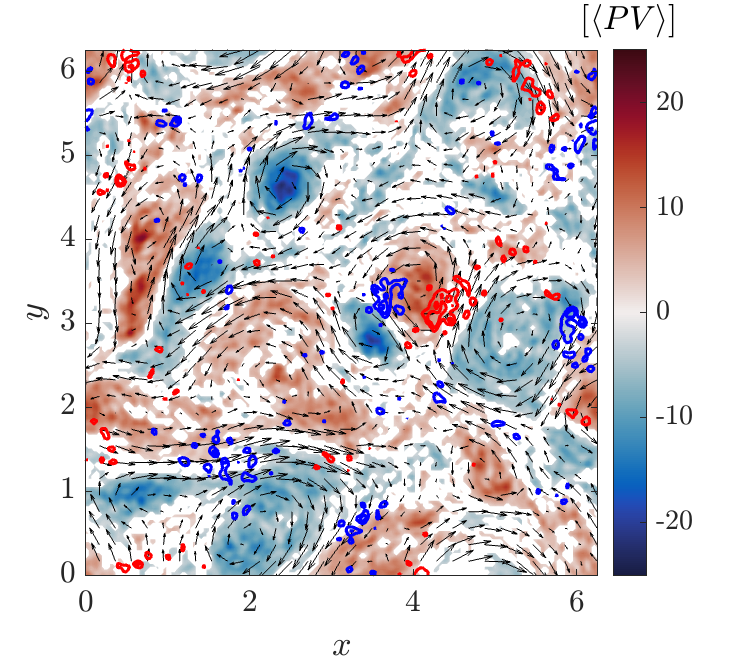}

    \caption{Column-integrated, time-averaged quantities for the no-rainfall case with $V_r=0$: Left: $[\langle H_s\rangle ]$ with contours of vertical velocity $[ \langle w\rangle ]$ in the ranges $ [ 0.1 W_{\rm max},  W_{\rm max}] \cup [ - W_{\rm max}, - 0.1 W_{\rm max}]$, with $W_{\rm max} = {\rm max}(|\langle w \rangle |) \approx 0.7$ and contour levels chosen as increments of $0.05 W_{\rm max}$ (red for positive, blue for negative). Right: $[\langle PV_s\rangle ]$ with horizontal velocity vectors $[\langle {\bf u}_h \rangle ] = [\langle u\rangle] \hat{\bf x}$ + $[\langle v\rangle ] \hat{\bf y}$ (black arrows). The contour levels in the bottom row are identical to those in the top row.}
    \label{fig:heightave-pv-h-w}
\end{figure}

\subsection{Column-integrated, time-averaged physical-space structures}
\label{sub-column}

As seen above, a large-scale structure of phase interfaces emerges from small-scale, random forcing because of nonlinear interactions.
Focusing on the no-rainfall case with $V_r=0$, here we explore how those interfaces are related to 
coherent updrafts/downdrafts represented by $\langle w\rangle$, and consequently, how they are also related to coherent structures of moist potential vorticity $\langle PV_s \rangle$. Since $PV_s$ is a slowly varying quantity, its variance will be small during the time averaging window $t\in [145,160]$.  Nevertheless, we apply the time average for consistency, and thus consider statistical relationships between $\langle H_s \rangle,$ $\langle w \rangle,$ and $\langle PV_s \rangle.$  
It is worth re-iterating that we are not studying the instantaneous vertical velocity \( w \), that is, the instantaneous updrafts and downdrafts. Instead, we are considering time-averaged waves \( \langle w \rangle \). This approach extracts the contribution from fast oscillations that, due to phase transitions, cannot be averaged out. The overarching question is: does the additional buoyancy nonlinearity due to phase transitions induce a distinct coupling between time-averaged vertical velocity and moist potential vorticity structures?
%arising from mechanism that is fundamentally different from dry dynamics?

To answer the above question, our first step is to visually establish the connection between the large-scale `fuzzy' phase boundaries as measured by $\langle H_s \rangle \approx 0.5$ (Figure \ref{fig:q_t_3d}), and the time-averaged vertical velocity $\langle w \rangle.$  
%This correlation is straightforward to establish, using column integration of $\langle H_s \rangle$ and $\langle w \rangle.$
In addition to the time average, we also apply a column integration to ensure
%The column-integration ensures 
a global examination of salient features, which may be %otherwise 
overlooked in single-layer evaluations. Denoting column integration by square brackets, we define
\begin{equation}
[ f ] (x,y,t) = \frac{1}{h} \int_{z_1}^{z_2} f(x,y,z,t) dz,
\label{eqn:column-integrated}
\end{equation}
where \( h = z_2 - z_1 \). For our purpose, we integrate across a vertical span centered at \( z=\pi \) and extending 12.5\% above and below, thereby vertically averaging over 25\% of the time-averaged data.  We are looking for structures that are coherent in time and vertical extent, but we do not demand vertical coherence over the entire height of the domain.  

The results for $[\langle H_s \rangle ]$ and $[\langle w \rangle ]$ are shown in Figure \ref{fig:heightave-pv-h-w} (top left), where one can see fuzzy phase boundaries in yellow.  The overlay of red and blue contours indicates locations where the magnitude of 
$[\langle w \rangle ]$ is greater than a threshold, in order to filter near-zero values. Specifically, we plot $[\langle w \rangle ] \in [ 0.1 W_{\rm max},  W_{\rm max}] \cup [ - W_{\rm max}, - 0.1 W_{\rm max}]$, where
$W_\textrm{max} = {\rm max}(|\langle w \rangle |)$ is the maximum absolute-value of $\langle w \rangle $ in the data set.  Positive contour levels are red and negative contour levels are blue. In a remarkable phenomenon, the red and blue structures are distributed along the yellow band-like structures, indicating that non-zero \( [ \langle w \rangle ] \) is linked to regions of frequent phase transitions with $[ \langle H_s \rangle ]\approx 0.5$. Indeed, a main effect of nonlinear buoyancy associated with phase interfaces is to reinforce coherent updrafts and downdrafts.  In Section 
\ref{sub-ODE}, this phenomenon will be further discussed in the context of an ODE model.

For comparison to dry dynamics, the bottom left panel of 
Figure \ref{fig:heightave-pv-h-w} shows the same contour levels of $[\langle w \rangle ]$ for the dry run with $Fr = Ro = \epsilon = 0.17$.  For this value $\epsilon$ approaching the upper cutoff to be considered as a small $(\epsilon < 1)$, the figure shows that updrafts and downdrafts exist, but they are less extensive than in the moist case. 
% Inspection of the higher $Fr$ case with $Ro = 0.17$ and $Fr = 0.24$ shows SOME STUFF (figure 1, tables, other figures).  Consistent with the spectrum, the analogous Figure 5 plots for $Ro = 0.17$, $Fr = 0.24$ are similar to Figure  5 bottom row, but with slihtly lower intensity vrotices, etc. MORE STUFF HERE.  Therefore there is a $Fr$-number effect.  However, we would like to point out that, in addition to the effect of the different $Fr$ numbers for unsaturated and saturated regions in the moist run, there is {\it also} an additional nonlinearity from the buoyancy term. 
% Analysis of the dry case with a slightly higher Froude number ($Ro = 0.17$, $Fr = 0.24$) reveals energy spectra (Figure~\ref{fig:energy_1}) and quantitative energy distributions (Table~\ref{table3}) closely resembling {(\color{red}some specific $x \%$)} those of the $Fr = Ro = 0.17$ case. 
Analysis of the dry case with a slightly higher Froude number ($Ro = 0.17$, $Fr = 0.24$) reveals energy spectra (Figure~\ref{fig:energy_1}) and quantitative energy distributions (Table~\ref{table3}, showing differences between the two cases in kinetic-to-total and potential-to-total energy ratios only at the first decimal place) closely resembling those of the $Fr = Ro = 0.17$ case.
Its flow structure is also similar to that shown in the bottom row of Figure~\ref{fig:heightave-pv-h-w} for $Fr = Ro = 0.17$, though with slightly weaker vortex intensity due to weaker inverse energy transfer
(not shown for conciseness; see also Sections~\ref{subsec:dynamicwaves} and \ref{subsec:NuNsimpact} for further quantitative comparisons). 

%Therefore, we omit a separate visualization for the $Fr = 0.24$ case in Figure~\ref{fig:heightave-pv-h-w}. { {\color{red} Notably, quantitative analysis of wave--potential vorticity interactions in Section~\ref{subsec:dynamicwaves} further confirms consistent results between these two dry cases. some simple words}} 

Comparing the moist case with the two dry baselines, we observe that the moist energy spectrum and inverse energy transfer fall between the two dry baselines. This reflects the varying effective $Fr$ (between 0.17 and 0.24) due to phase changes, suggesting a potential bridge between moist and dry dynamics. More importantly, however, the piecewise-defined buoyancy introduces additional nonlinearity, generating extra nonlinear waves and more complex wave--potential vorticity interactions. As a result, the relationship between dry and moist dynamics is not a simple superposition, complicating rigorous theoretical analysis. In Sections~\ref{subsec:dynamicwaves} and \ref{sub-ODE}, we will further elucidate the underlying nonlinear interaction mechanisms through quantitative analysis and theoretical derivation.

Fast-wave averaging theory suggests that the dry updrafts and downdrafts will be diminished as $Ro=Fr=$ $\epsilon$ is decreased. Indeed, for smaller parameter values ($Ro \approx 0.017$, $Fr_u \approx 0.017$, and $Fr_s \approx 0.024$), this is verified in Section \ref{subsec:dynamicwaves} and Appendix \ref{subsec: smaller parameters}. In Section \ref{subsec:dynamicwaves}, we quantitatively assess the wave–potential vorticity interactions, showing that it is approximately 19 times stronger in the moist case compared with the dry case. In Appendix \ref{subsec: smaller parameters}, we provide a qualitative illustration of the dynamics by presenting the analog of Figure \ref{fig:heightave-pv-h-w} (see the bottom left panel of Figure \ref{fig: heightave-pv-h-w, appA}).
Figure \ref{fig: heightave-pv-h-w, appA} (top left) shows more extensive drafts in the case of phase changes than without them, though we remind the reader that fast-wave-averaging theory remains open for the case with phase changes.

% {\color{red}Indeed, we verify this in section 4.6 and appendix A. in section 4.6 we quantitatively calculate wave-potential voriticity interaction, 19 times stronger when comparing dry and moist. in appendix A we visualize the dynamics via presenting the analog of figure 5  }

% Indeed we verify this in Appendix A, where we present the analog of Figure \ref{fig:heightave-pv-h-w} for smaller $Ro \approx 0.017$, $Fr_u \approx 0.017$ and $Fr_s \approx 0.024$ (see the bottom left panel of Figure \ref{fig: heightave-pv-h-w, appA}). Figure \ref{fig: heightave-pv-h-w, appA} (top left) shows more extensive drafts in the case of phase changes than without them, though we remind the reader that fast-wave-averaging theory remains open for the case with phase changes.

Next we turn attention to the column-integrated, time-averaged potential vorticity. Figure 
\ref{fig:heightave-pv-h-w} (right) shows that moist $[\langle PV_s\rangle]$ (top) and dry $[\langle PV \rangle]$ (bottom) are localized within positive (red) and negative (blue) structures. Furthermore, by overlaying horizontal velocity vectors $[\langle {\bf u}_h \rangle ] = [\langle u\rangle] \hat{\bf x}$ + $[\langle v\rangle ] \hat{\bf y}$, we conclude that these positive and negative structures are, respectively, cyclonic and anticyclonic vortices.  

%It also appears that these $PV_s$-vortices % are roughly aligned
%\begin{color}{red} have a connection \end{color}
%with the coherent updrafts/downdrafts, which will be quantitatively explored below. 

Focusing on the moist simulation, Figure \ref{fig:heightave-pv-h-w} (top row), one can see different types of structures.  Inside sub-box 1 (near $x=4.5,y=4.5$), drawn in the upper right corner of the top right panel, there is a region of downdrafts in between counter-rotating $PV_s$-vortices.  In this `dipole,' the positive $PV_s$-vortex (cyclone) is stronger and more coherent that the negative $PV_s$ region (anticyclone). Note that Figure \ref{fig:heightave-pv-h-w} also shows strong updraft regions in between positive and negative $PV_s$-structures, such as near $(x=1.8,y=2.8)$ and near $(x=4.7,y=2.5)$.
A second sub-box 2 (near $x=3,y=0.5$) encloses a  $PV_s$-anticyclone flanked on either side by an updraft and a downdraft. Sub-box 1 and 
sub-box 2 are representative structures in the flow, and their 3D structure is visualized in Appendix \ref{subsec:3D_HwPV_s}.

%We have chosen sub-box 1 and sub-box 2 as representative structures in the flow, and visualize their 3D structure in {\color{blue}Appendix \ref{subsec:3D_HwPV_s}}.  
% We have chosen sub-box 1 and sub-box 2 as representative structures in the flow, and visualize their 3D structure in the next section.  

Altogether, the information in Figure \ref{fig:heightave-pv-h-w} gives an overview of the large-scale, slowly varying structures that have developed spontaneously from small-scale, randomly-forced fluctuations.  When phase changes are present, Figure \ref{fig:heightave-pv-h-w} (top) confirms that nonlinear buoyancy plays an important role by 
boosting $\langle w \rangle$ near
phase interfaces $\langle H_s \rangle$.  It is also evident that 
these coherent structures $\langle w \rangle$ and $\langle H_s \rangle$ are connected in time and space with slowly varying moist potential vorticity $PV_s$.

\subsection{Statistical analyses of coherent structures in flow regions with $PV_s$-vortices}
\label{subsec:statistical_analysis}

It is evident that $PV_s$-vortices, phase interface bands, and regions of high vertical velocity exhibit a significant degree of spatial overlap, underscoring their interconnected dynamics (see also Appendix \ref{subsec:3D_HwPV_s}).
Here we investigate statistical relationships between $\langle H_s\rangle$, $\langle w\rangle$ and $\langle PV_s\rangle$.
% (general)

% (why global not working, and therefore we turn to local.)

% Intuitively, we first consider conducting a statistical analysis of \( \langle w \rangle \) from a global perspective. However, after computation (not shown here), we find that the probability density function of \( \langle w \rangle \) approximately follows a Gaussian distribution with zero mean, which offers little additional insight. Therefore, we shift our focus to the statistics of \( \langle w \rangle \) over local regions.

Intuitively, we first consider a global statistical analysis of $\langle w \rangle$. It should be noted that the white-noise forcing applied to all fields ($u$, $v$, $w$, $\theta_e$, $q_t$), combined with the periodic boundary conditions suggests Gaussian statistics for global quantities of interest such as $\langle w \rangle$ and $\langle q_t \rangle$. This expectation is confirmed by numerical computations (not shown here), which demonstrate that the probability density functions of $\langle w \rangle$, $\langle q_t\rangle$ and $\langle PV_s \rangle$
closely follow a zero-mean Gaussian distribution. Since this global statistical result provides limited additional physical insight, we shift our focus to the statistical properties in local regions. To further explore the relationships among \( \langle H_s \rangle \), \( \langle w \rangle \), and $\langle PV_s \rangle$, this study first examines local areas dominated by relatively high $PV$ values---specifically, regions that predominantly contain clear cyclonic or anticyclonic \( PV_s \)-vortices. We analyze the statistical behaviors of \( \langle w \rangle \) and \( \langle H_s \rangle \) over these regions to investigate their quantitative linkages. Accordingly, \( \langle w \rangle \) is conditioned on \( \langle PV_s \rangle \) for the moist case and on \( \langle PV \rangle \) for the dry case (see Figure~\ref{fig:heightave-pv-h-w}), to better capture its relationship with vertical velocity.

In this section, we condition the main moist simulation on \( \langle PV_s \rangle \), which means our analysis focuses on a subset of statistics from regions with $PV_s$-vortices. This region selection is particularly advantageous because it lends itself to straightforward automation and is relevant for applications \citep{chagnongray09,weijenborgetal2015,weijenborgetal2017}. Furthermore, results will be compared with dry runs in Section \ref{subsec:NuNsimpact} when assessing robustness to $R_{fr}$. Other region types—such as phase interface bands or regions with high vertical velocity—could also be used for future analyses. 

\subsubsection{Region selection for statistical analyses}
\label{subsub:region_selection_sta}

The $PV_s$-vortical regions are identified using the following method.
%\textbf{Filtering Methodology:} 
The \( \langle PV_s \rangle \) data is filtered according to the formula
\begin{equation}
\{ PV_s \,|\, PV_s \notin (\mu \pm a\sigma) \} 
\label{eqn:pv-filter}
\end{equation}
where the probability density function of \( PV_s \) is approximately Gaussian with mean value \( \mu \approx 0 \) and standard deviation \( \sigma \approx 10.5 \). The parameter \( a \) is a multiplicative factor used to capture high-magnitude information. In practice, we use \( a = 1.7 \), corresponding to \( |PV_s| > 17.9 \), to ensure good coverage of the cyclonic/anticyclonic structures while retaining enough data for statistically meaningful results. If \( a \) is too small (e.g., \( a = 1 \)), then the coherent structures are not sufficiently isolated. On the other hand, if \( a \) is too large (e.g., \( a = 2 \)), then the sample size becomes too small for further analysis. 

Simultaneously, the corresponding \( \langle w \rangle \) and \( \langle H_s \rangle \) data are automatically filtered by association with the \( \langle PV_s \rangle \) data. Once the locations of the $PV_s$-vortices have been determined by the filtering, 
%the filtered \( \langle PV_s \rangle \) data is determined, focusing on the \( PV_s \)-vortices, 
then the \( \langle w \rangle \) and \( \langle H_s \rangle \) data within these regions are also collected.

\subsubsection{Vertical velocity and phase interfaces within $PV_s$-vortices}

The filtered data is used to investigate the statistical relationship between phase interfaces \( \langle H_s \rangle \) and vertical velocity \( \langle w \rangle \) within $PV_s$-vortices. This analysis is conducted by examining the kurtosis (fourth-order moment) of the filtered vertical velocity data \( \langle w \rangle \) and the probability density functions (PDFs) of the filtered phase-interface data \( \langle H_s \rangle \). 
%The filtered regions, corresponding to coherent cyclonic/anticyclonic \( PV_s \)-vortices, serve as a natural framework for studying how phase interfaces interact with vertical velocity structures.

%The choice to analyze the kurtosis of \( \langle w \rangle \) is motivated by the work of \textcolor{red}{Marino et al. (Rev. Fluids 7, 033801, 2022)}, which demonstrated 
For dry, stably stratified dynamics, \cite{marinoetal2022} demonstrated that intermittent, strong vertical drafts at large scales, reflected by high kurtosis values, are connected to high turbulence dissipation events. Their work supports evidence that the ocean dissipation occurs mainly in localized regions of enhanced mixing. Following a similar methodology as \cite{marinoetal2022}, here we investigate the connection between phase interfaces and intermittent, large-scale, vertical drafts within $PV_s$-vortices, with possible relevance to storm dynamics in atmospheric flows.
%As kurtosis reflects the intermittency of the flow, this analysis investigates whether regions with high intermittency in vertical drafts influence the distribution and frequency of phase transitions. Such an approach provides statistical insights into the interaction and coupling between phase interfaces and vertical velocity.

\begin{figure}
    \centering
    \includegraphics[width=0.4\textwidth]{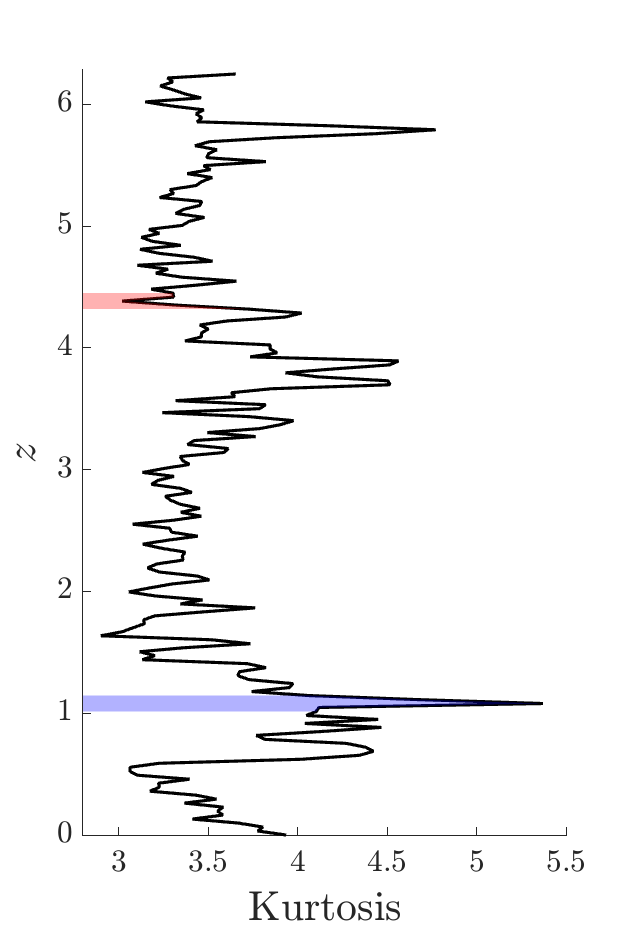}
    \includegraphics[width=0.5\textwidth]{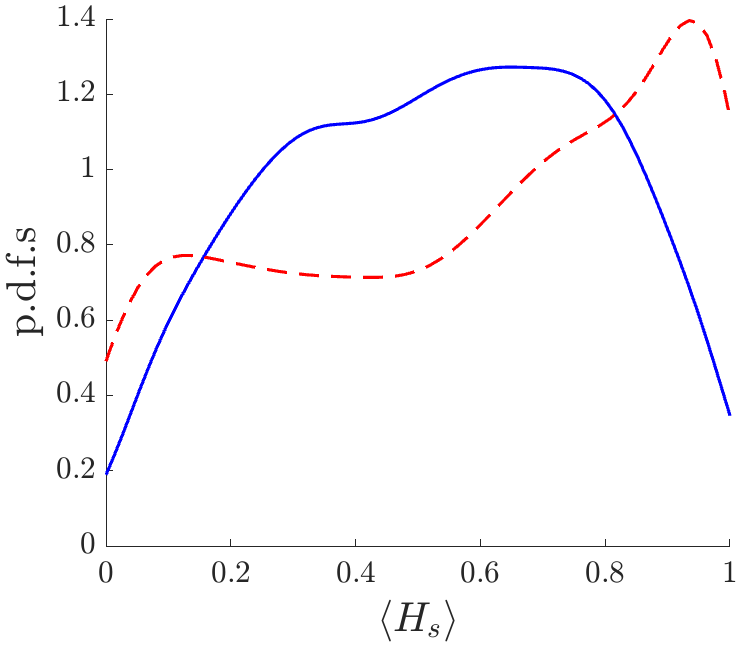}
    \caption{Left: Kurtosis of filtered vertical velocity \( \langle w \rangle \) as a function of the vertical coordinate \( z \). The kurtosis profile highlights regions of high intermittency (peaks) and low intermittency (troughs). Right: Probability density functions (PDFs) of filtered phase boundary \( \langle H_s \rangle \) corresponding to selected high-kurtosis (left purple region) and low-kurtosis $z$-intervals (left pink region). The PDF of $\langle H_s \rangle$ for the high-kurtosis region (solid curve) exhibits values concentrated in the central range \( (0.3, 0.7) \), indicating frequent phase transitions.  The PDF of $\langle H_s \rangle$ for the low-kurtosis region (dashed curve) shows higher values \( (0.0, 0.1) \) and \( (0.9, 1.0) \), indicating less frequent phase transitions.}
    \label{fig: kurtosis and pdfs}
\end{figure}

%\textbf{Kurtosis of \( \langle w \rangle \)}

For each height $z$ in the filtered 3D data set, we compute the kurtosis of $\langle w \rangle$ given by 

\begin{equation}
{\rm kurtosis}(\langle w \rangle ) = \frac{\int_A (\nabla^2 \langle w \rangle )^4 dA}
{\biggl ( \int_A (\nabla^2 \langle w \rangle )^2 dA\biggr )^2},
\end{equation}
where $A$ is the area of a horizontal slice.
The left panel of Figure~\ref{fig: kurtosis and pdfs} shows that the kurtosis varies in the range $(2.7,5.5)$, where peaks in the profile correspond to high intermittency, while troughs correspond to low intermittency. Two representative \( z \)-intervals are selected for further analysis: a high-kurtosis region (marked in light purple), and a low-kurtosis region (marked in light pink). These intervals provide a basis for examining the relationship between $\langle w \rangle$-intermittency and phase transitions.

%\textbf{PDFs of \( \langle H_s \rangle \)}

The right panel of Figure~\ref{fig: kurtosis and pdfs} displays the PDFs of the filtered phase-interface data \(\langle H_s \rangle \) corresponding to the selected high-kurtosis and low-kurtosis regions. The distributions reveal distinct characteristics.  In the high-kurtosis region (purple), the \( \langle H_s \rangle \) values are predominantly distributed in the central range \( (0.3, 0.7) \), indicating frequent phase transitions. 
 In contrast, the low-kurtosis region (pink) has \( \langle H_s \rangle \) values accumulating near the boundaries, within the intervals \( (0.0, 0.1) \) and \( (0.9, 1.0) \), in accordance with less frequent phase transitions.  These results are suggestive of two-way feedback between coherent updrafts/downdrafts and large-scale phase boundaries, which is a topic for future research.

\subsubsection{$PV_s$-vortices and vertical velocity}
\label{subsec:statwPV_s}

The filter described in \eqref{eqn:pv-filter} is applied to the column-integrated data \([\langle PV_s \rangle]\), where the probability density function of \([\langle PV_s \rangle]\) is approximately Gaussian with mean value \(\mu \approx -0.2\) and standard deviation \( \sigma \approx 4.9 \). For the value \(a=1.7\), the filtered data corresponds to \(|PV_s|>8\).  As before, the corresponding \( [\langle w \rangle] \) data is automatically filtered along with the \( [\langle PV_s \rangle] \) data. We may then proceed to investigate the statistical relationship between \( [\langle PV_s \rangle] \) and \( [\langle w \rangle] \) within the filtered, column-integrated, time-averaged data. 
For simplicity, the focus is on magnitudes 
\( |[\langle PV_s \rangle]| \) and \( |[\langle w \rangle]| \)
only, leaving the more detailed study of signed information for future work aimed at targeted geophysical scenarios.

For ease of notation, we denote the data set by \( (PV_s, w)_f \), which is 
%The filtered dataset \( (PV_s, w)_f \) is 
further divided into quartiles based on the magnitude \( |[\langle PV_s \rangle]| \). The first quartile \( (PV_s, w)_1 \) represents the lowest 25\% of \( |[\langle PV_s \rangle]| \) values, and the subsequent quartiles \( (PV_s, w)_2, (PV_s, w)_3, (PV_s, w)_4 \) contain progressively higher magnitudes \( |[\langle PV_s \rangle]| \), with the fourth quartile capturing the highest 25\%. This grouping allows us to systematically investigate the trends in \(  [\langle w \rangle]  \) associated with increasing magnitudes \( |[\langle PV_s \rangle]| \). Figure~\ref{fig:l2norm-w} shows the spatial distribution of the \( |[\langle PV_s \rangle]| \)-quartiles over the domain. Different symbols (\( \times, *, \circ, \triangle \)) denote the quartiles 0-25\%, 25-50\%, 50-75\%, and 75-100\%, respectively. Most data points from the highest quartiles are located within the cores of \( PV_s \)-vortices.
%confirming their strong association with coherent cyclonic and anticyclonic structures.

For each quartile, the normalized \( L_2 \)-norm of \( [ \langle w \rangle ]\) is computed. The normalization factor is the \( L_2 \)-norm of \( [\langle w \rangle ] \) in the first quartile \( (PV_s, w)_1 \), which serves as a baseline for comparison. The results in Table~\ref{table: l2norm-w} show a clear trend: the normalized \( L_2 \)-norm of \( [\langle w \rangle ]\) increases with \( |[\langle PV_s \rangle]| \). Specifically, the highest quartile \( (PV_s, w)_4 \) exhibits a 27.7\% increase in the \( L_2 \)-norm of \( [\langle w \rangle ] \) compared to the first quartile. 
Figure~\ref{fig:l2norm-w} and Table~\ref{table: l2norm-w} demonstrate that higher magnitudes $ | [ \langle PV_s \rangle ] |$ and higher magnitudes $ |[\langle w \rangle ] |$ are statistically linked within $PV_s$-vortices.

\begin{figure}%[H]
\centering
\includegraphics[width=0.5\textwidth]{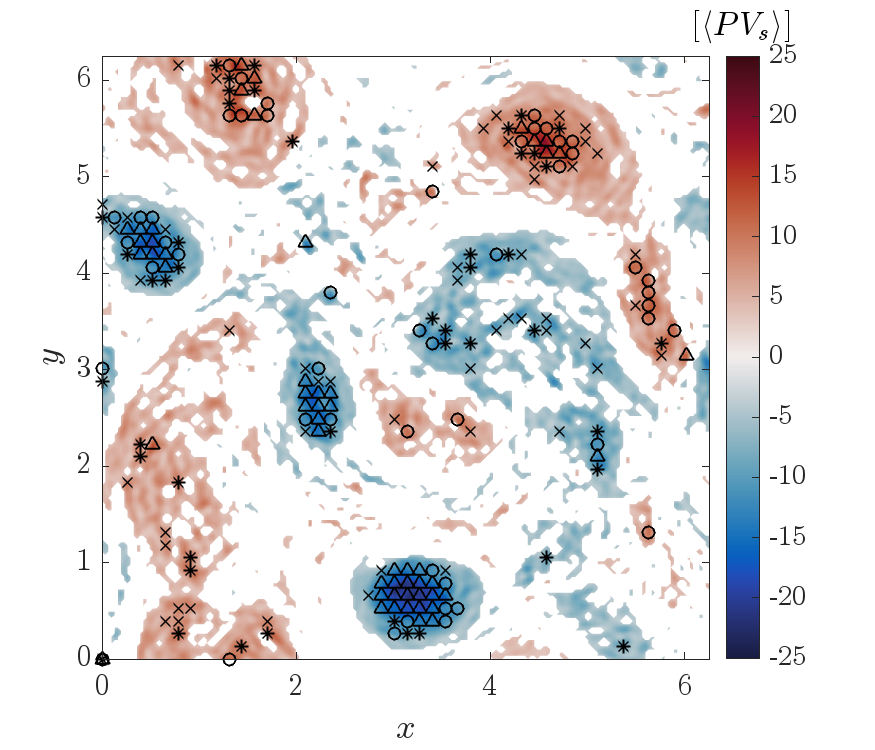}
\caption{Column-integrated, time-averaged moist potential vorticity $[\langle PV_s\rangle]$. Four different symbols are used to highlight the $| [ \langle PV_s \rangle ] |$ data beyond 1.7 standard deviations, where $\times,\ *,\ \circ,\ \triangle$ correspond to,
respectively, the 0-25\%, 25\%-50\%, 50\%-75\% and 75\%-100\% quartiles of filtered $ | [\langle PV_s \rangle ] |$ data.
%Right: normalized $L_2$ norm of $\langle w\rangle$. The four points marked with * show the $L_2$ norm values of $w$ of datasets $(PV, w)_1$ - $(PV, w)_4$ respectively, divided by the $L_2$ norm of $w$ of $(PV, w)_1$.}
\label{fig:l2norm-w}
}
\end{figure}

\begin{table}
    \centering
    \begin{tabular}{cccc}
        \hline
          $(w_{L_2})_1$ &  $(w_{L_2})_2$ &  $(w_{L_2})_3$ &  $(w_{L_2})_4$ \\
        \hline
         1 & 1.0436 & 1.0935 & 1.2771 \\
        \hline
    \end{tabular}
\caption{Normalized $L_2$ norm of $[\langle w\rangle]$ in datasets $(PV_s, w)_1$--$(PV_s, w)_4$.}
\label{table: l2norm-w}
\end{table}

\subsection{Dynamical coupling between $PV_s$ and time-averaged waves}
\label{subsec:dynamicwaves}

 Here we aim to assess coupling between fast waves and slowly varying $PV_s$, {\it i.e}, wave-vortical coupling. To do so, we consider the evolution equation for $PV_s$ and use the nonlinear wave-vortical decomposition 
\citep{ss17,zhang2021effects,zhang2021fast,zhang2022convergence,remondtiedrez-decomposition2024}.  Following the decomposition, a scalar variable $f$ is written as $f = f_\textrm{(pv,m)} + f_\textrm{(wave)},$ where $f_\textrm{(pv,m)}$ is the slow part found using $PV$-and-$M$ inversion, and $f_\textrm{(wave)}$ is the fast-wave part (see Appendix \ref{subsec:slow-fast-coupling}).

\begin{comment}
{\color{blue}Using the decomposition described in Appendix \ref{subsec:slow-fast-coupling}, one may analyze the evolution equation for slowly varying, moist potential vorticity $PV_s$, as well as its time-averaged version.  In the limit $Fr_s \sim Fr_u \sim Ro = \epsilon \rightarrow 0$, the time-averaged $PV_s$-equation is referred to as the `fast-wave-averaged' equation. Pertaining to our simulations, we are interested in finite $\epsilon$ small, and we aim to quantitatively assess coupling between $PV_s$ and time-averaged waves on $O(1)$ time scales \citep{zhang2021effects,zhang2022convergence}.
As in previous sections, we consider the simulation with $V_r=0, N=192$ during the time window $t \in [145,160]$, which is approximately the last 10\% of the simulation.}
\end{comment}

The evolution equation for $PV_s$ may be derived by combining the curl of the horizontal momentum equation with the partial derivative with respect to $z$ of the equation for the saturated buoyancy.  It may be written as 
\begin{equation}
\frac{\partial PV_s}{\partial t} = -  {\bf u}\cdot \nabla PV_s  - {\cal{F}}\frac{\partial {\bf u}}{\partial z} \cdot \nabla b_s - \xi \biggl (\frac{\partial u}{\partial x} + 
\frac{\partial v}{\partial y}\biggr ) - 
\biggl (\frac{\partial w}{\partial x}\frac{\partial v}{\partial z} - 
\frac{\partial w}{\partial y}\frac{\partial u}{\partial z}\biggr ),
\label{eqn:PV_s-original}
\end{equation}
where $\xi = \hat{\bf z} \cdot \nabla \times {\bf u}$ is the vertical component of vorticity \citep{zhang2021fast}. 
Equation \eqref{eqn:PV_s-original} is the Boussinesq counterpart of the $PV_s$-equation found in the quasi-geostrophic limit $Fr_s \sim Fr_u \sim Ro =\epsilon \rightarrow 0$ \citep{ss17}.
The limiting equation has the following modifications from \eqref{eqn:PV_s-original}: the velocity vector ${\bf u}$ is replaced by ${\bf u}_h$ in the first two terms on the right-hand-side; the second `phase-change term' is identically zero in saturated regions and is only activated when the environment is locally unsaturated; and the nonlinear terms vanish.

The time average \eqref{eqn:bracketavg} may be applied to each term on the right-hand-side of \eqref{eqn:PV_s-original}. 
In lieu of a full budget analysis, we focus on the time-averaged advection term $\langle {\bf u} \cdot \nabla PV_s \rangle,$ to assess the relative contributions arising from ${\bf u}_\textrm{(pv,m)}$ and ${\bf u}_\textrm{(wave)}$.
Since $PV_s$ is a slowly varying quantity, we make the approximation
\begin{equation}
\langle {\bf u} \cdot \nabla PV_s \rangle \approx 
\langle {\bf u} \rangle \cdot \langle \nabla PV_s \rangle.
\label{eqn:advectionapprox-1}
\end{equation}
Then using the decomposition \eqref{def:slowwave}, we may rewrite \eqref{eqn:advectionapprox-1} as
\begin{equation}
\langle {\bf u} \cdot \nabla PV_s \rangle \approx 
\langle {\bf u}_\textrm{(pv,m)}\rangle \cdot \langle \nabla  PV_s \rangle + 
\langle {\bf u}_\textrm{(wave)}\rangle \cdot \langle \nabla PV_s \rangle,
\label{eqn:advectionapprox-2}
\end{equation}
and consider the ratio of the terms on the right-hand-side of \eqref{eqn:advectionapprox-2}, anticipating that phase changes will make the latter term more important than in dry dynamics.

Before comparison of time-averaged terms in \eqref{eqn:advectionapprox-2} may be considered, we must first decompose the instantaneous velocity fields ${\bf u}$ into their $\textrm{(pv,m)}$ and $\textrm{(wave)}$ parts.
To find ${\bf u}_\textrm{(pv,m)}$ from the numerical data at any fixed time, we use an iterative procedure, where the nonlinear elliptic equation \eqref{PVsMinversion} in $PV_s$-and-$M$ inversion is approximated by 
\begin{equation}
\nabla_h^2\psi+ \dfrac{\partial}{\partial z} \biggl [ \dfrac{1}{1+{\cal G}}({\cal F}^2\partial_z\psi + {\cal F}M) H_u + ({\cal F}^2\partial_z\psi)H_s \biggr ] \approx PV_s.
\label{PVsMinversion-approx}
\end{equation}
with corresponding approximation for the hydrostatic balance relation
\begin{equation}
\theta_e \approx \frac{1}{1+{\cal G}}\biggl ( {{\cal F}}\partial_z \psi + M \biggr ) H_u + \biggl ({{\cal F}}\partial_z \psi \biggr ) H_s.
\label{inversion-theta-approx}
\end{equation}
In \eqref{PVsMinversion-approx} and \eqref{inversion-theta-approx}, the Heaviside functions $H_u(q_t), H_s(q_t)$ are calculated from the data, and they are mixed fields containing both $\textrm{(pv,m)}$  and $\textrm{(wave)}$ information. Therefore the solution $\psi$ is not a purely slow pressure, but rather is contaminated by waves.  However, approximate $PV_s$-and-$M$ inversion using \eqref{PVsMinversion-approx}-\eqref{inversion-theta-approx}, together with the definition \eqref{def:PV_s} for $PV_s$, may be iterated until convergence.  In just a few iterations, the process converges to the solution of \eqref{PVsMinversion}-\eqref{balance-relations}.

We calculate ${\bf u}_\textrm{(pv,m)}$ and ${\bf u}_\textrm{(wave)}$ for each time $t \in [145,160]$, and then find the averages $\langle {\bf u}_\textrm{(pv,m)} \rangle$ and $\langle {\bf u}_\textrm{(wave)} \rangle$.
Finally, we compute the normalized ratio $R_\textrm{avg}/R_0$ given by
\begin{equation}\frac{R_\textrm{avg}}{R_0} = \frac{1}{R_0}
\dfrac{\Vert \langle {\bf u}_\textrm{(wave)}\rangle \cdot \langle \nabla PV_s \rangle \Vert_{L_2}}{\Vert \langle {\bf u}_\textrm{(pv,m)}\rangle \cdot \langle \nabla PV_s\rangle \Vert_{L_2}}, \quad 
R_0 = \dfrac{\Vert {\bf u}_\textrm{(wave)}  \cdot  \nabla PV_s \Vert_{L_2}}{\Vert {\bf u}_\textrm{(pv,m)}  \cdot \nabla PV_s \Vert_{L_2}}\; {\Big \rvert_{t=t_0}},
\label{Ravgnorm}
\end{equation}
where $t_0=145$ is that start time for the averaging window. 
 In the two dry simulations with $Ro=Fr=0.17$ and $Ro=0.17, Fr=0.24$, we find $R_\textrm{avg}/R_0 \approx 0.19$, whereas in the phase-change simulation we find a $R_\textrm{avg}/R_0 = 0.43$, a factor of roughly 2.26 higher. This factor is notable, in light of 
%\begin{color}{red} finite $\epsilon = O(0.1)$ \end{color}, and 
the fact that only regions of high-frequency phase changes with $\langle H_s \rangle \approx 0.5$ contribute to differences between $\langle {\bf u}_\textrm{(wave)}\rangle$ in the dry and phase-change cases (see Figure \ref{fig:heightave-pv-h-w}). On the other hand, the $L_2$ norm is global and thus includes data from large portions of the domain which are purely saturated or purely unsaturated during the time window $t\in [145,160]$. Moreover, when moving to a smaller parameter regime 
($Ro = Fr_u = 0.017, \; Fr_s = 0.024$), the enhancement becomes much stronger: 
we find $R_{\text{avg}}/R_0 \approx 0.664$, which is nearly 19 times larger 
than the corresponding dry value of $0.035$ at $Ro = Fr = 0.017$. For the term $\langle {\bf u} \cdot \nabla PV_s \rangle$, we conclude that the presence of phase changes significantly increases coupling between waves and moist potential vorticity, as expected.
% \begin{color}{red}Should we add the 2nd dry case?\end{color}

\subsection{Nonlinear waves in an ODE system with phase changes}
\label{sub-ODE}

Here we present a model of coupled ordinary differential equations to illustrate the nature of nonlinear waves that arise in the presence of a phase boundary. The model was first discussed in \cite{zhang2021effects}, and is revisited here so that results of this manuscript are self-contained; it serves to explain the time-average vertical velocity \(\langle w \rangle\) and coupling term \(\langle {\bf u}_\textrm{(wave)}  \rangle \cdot \langle \nabla PV_s \rangle\) discussed above.

Eliminating spatial variations, the ODE model describes fast oscillations of vertical velocity $w$ and buoyancy $b$ at a fixed point in space, and is given by
\begin{equation}
    \frac{dw}{dt} = b = N_ub_u H_u + N_sb_s H_s,
    \quad
    \frac{db_u}{dt} + N_u w = 0,
    \quad
    \frac{db_s}{dt} + N_s w = 0, 
    \label{eqn:ode-system-condensed}
\end{equation}
where the waves represented by \eqref{eqn:ode-system-condensed} have different frequencies $(N_u,N_s)$ in different phases $(b_u,b_s)$.  The phase boundary is determined by the condition $b_u=b_s$ and the cloud indicator $H_s({\bf x},t)$ may be written as $H_s(b_s - b_u)$~\citep{marsico2019energy}. Noticing that the quantity $M = N_u^{-1} b_u - N_s^{-1} b_s$ is an invariant of \eqref{eqn:ode-system-condensed}, we may rewrite the system with the replacement of $b_s$ as 
\begin{align}
    &\frac{dw}{dt} = N_u b_u H_u + N_s\left(\frac{N_s}{N_u}b_u - M N_s\right) H_s, \\
    &\frac{db_u}{dt} + N_u w = 0.
\end{align}
Eliminating $w$ leads to a nonlinear oscillator equation for $b_u$ with piecewise general solution given by
\begin{equation}
    b_u = \begin{cases}
        c_{u1}\sin(N_ut) + c_{u2}\cos(N_ut), & \text{for unsaturated regions}, \\
        c_{s1}\sin(N_st) + c_{s2}\cos(N_st) + MN_u, & \text{for saturated regions}.
    \end{cases}
    \label{eqn:soln-ode}
\end{equation}

For the special case $M=0$, we can evaluate the coefficients in \eqref{eqn:soln-ode} starting from the initial conditions \( b_u(t_0) = 0 \) and \( w(t_0)=a\) with \( a > 0 \). As shown in figure 13 of \cite{zhang2021effects}, the solution first enters the saturated phase with \( b'_u(t_0) = -N_u a < 0 \).  It is a piecewise sine function, alternating between frequencies \( N_u \) and \( N_s \), with period \( \mathcal{T} = \pi N_u^{-1} + \pi N_s^{-1} \), such that
\begin{gather}
b_u(t) =  \left\lbrace \begin{aligned}
& -\dfrac{aN_u}{N_s} \sin(N_s(t - t_0)) \hspace{0.3cm} t \in [t_0 + n {\cal{T}}, t_0 + \dfrac{\pi}{N_s} + n{\cal{T}}]   \\
%&\hspace{7cm}\text{where } {\cal{T}} = \dfrac{\pi}{N_u} + \dfrac{\pi}{N_s} \text{ and } n = 0, 1, 2 \cdots\\
& a \sin(N_u(t - t_0 - \dfrac{\pi}{N_s})) \hspace{0.30cm} t \in [t_0 + \dfrac{\pi}{N_s} + n{\cal{T}}, t_0 + (n+1){\cal{T}}],
\end{aligned} \right. 
\label{soln:b_u}
\end{gather}
where $n = 0, 1, 2 \cdots$.  The nonlinear wave solution \eqref{soln:b_u} is qualitatively the same as the one illustrated in figure 13 of \cite{zhang2021effects}.

For \( a = 1 \) and \( N_u/N_s = R_{fr} \), we can examine the time-averaged value of \( b_u(t) \) in cases with and without changes of phase.  We are particularly interested in the strongly stratified regime with relatively large $N_u \sim N_s$, as in our numerical computations and some instances of stably stratified flows in nature.  For illustrative purposes, it is then helpful to consider the fast-wave-averaging limiting $N_u \rightarrow \infty$ with 
$R_{fr} = O(1)$.

% \begin{figure}%[H]
%     \centering
%     \includegraphics[width=.8\textwidth]{fig/Bu_1/odefigure.png}
%     \caption{Sketch of the nonlinear wave solution \eqref{soln:b_u} with nonzero time average.}
%  \label{fig:ode}
% \end{figure}

In the purely unsaturated case\footnote{This case is equivalent to dry dynamics.} with $R_{fr}=1$, the analytical solution \eqref{soln:b_u} for $b_u(t)$ is a simple sine function with frequency $N_u$. In that case, and for fixed averaging interval ${T}$, the average $ | \frac{1}{T} \int_0^{T} b_u(t) dt | \leq \dfrac{2}{N_uT} \rightarrow 0$ as $N_u \rightarrow \infty $. However, when the phase boundary is present, the parameter $R_{fr}$ is strictly greater than one. Then the time average is bounded by
\begin{equation}
\frac{2a}{\pi}\left(R_{fr}-1\right) - \frac{2a}{N_uT} \leq \left|\frac{1}{T}\int_0^Tb_u(t)\text{d}t\right| \leq \frac{2a}{\pi}\left(R_{fr}-1\right) + \frac{2a}{N_uT}.
\label{eq:timeaveint-R}
\end{equation}
For $N_u \rightarrow \infty$, $R_{fr} > 1,$ the time average achieves the nonzero value $2a \pi^{-1} (R_{fr}-1)$, dependent on $R_{fr}$.  In the next Section \ref{subsec:NuNsimpact}, we investigate the impact of $R_{fr} = N_u/N_s$ on the time-averaged quantities $\langle H_s\rangle$ and $\langle w \rangle$ in our numerical simulations.

% \begin{color}{red}  Talk about the left of figure 11 for the case $R_{fr}=\sqrt{2}.$
% \end{color}

\subsection{Dependence of coherent structures on \( R_{fr}=N_u/N_s \)}
\label{subsec:NuNsimpact}

Compared to dry Boussinesq system, fundamentally new physics in our moist model \eqref{eqn:boussinesq} is the difference in wave propagation frequencies for the vapor and liquid states. This difference is incorporated into the nonlinear buoyancy defined by \eqref{eqn:buoyancy}-(\ref{eqn:bu-bs-def}) with distinct buoyancy frequencies \( N_u \), \( N_s \) given by (\ref{eqn:Nu-Ns-def}). Furthermore, the setup of the ODE model (\ref{eqn:ode-system-condensed}) and its exact solution \eqref{soln:b_u} demonstrate the piecewise nature of the nonlinear waves generated by the nonlinear buoyancy. In the context of the ODE model, the inequality (\ref{eq:timeaveint-R}) shows that the time average of the exact solution \( \langle b_u \rangle \) (and similarly $\langle b_s \rangle $) is bounded away from zero, and that its fast-wave-averaging limit is a nonzero value that depends on the  ratio \( R_{fr} = N_u/N_s \) of buoyancy frequencies. In the context of the moist Boussinesq system, such nonzero time-averaged waves induce nonlinear coupling in $PV_s$-evolution equation~(\ref{eqn:PV_s-original}), thereby affecting energy transfer and coherent structures that develop in the flow.

It is natural, then, to consider the influence of $R_{fr} = N_u/N_s$ on the formation and evolution of coherent structures, as well its influence on the statistics associated with the coherent structures.  The ratio $R_{fr}=1$ is the degenerate case without phase transitions, and hence we
vary \( R_{fr} \) in the range $R_{fr} \in (1, 3]$.  The results in previous sections set $R_{fr} = \sqrt{2}$, and we require values $R^2_{fr} < 10$ to ensure positivity in the energy equation~(\ref{eq:energy-vt}).  Here we focus on $R_{fr} \leq 3$ in order to remain close to the parameter regime $Fr_s \sim Fr_u \sim Ro = \epsilon \ll 1$ associated with the quasi-geostrophic regime (Table \ref{table: parameter setting}). For our simulations, the ratio $R_{fr}=3$ implies $Fr_s \approx 0.5$, which is borderline in the QG regime.

Starting with an examination of total energy spectra, Figure \ref{fig:spectrumNuNs} shows that the case $R_{fr} = 1.1$ accumulates the most energy, and the case $R_{fr}=3$ accumulates the least energy. 
There is a continuous decrease in total energy for values $R_{fr} \in (1, 3]$. As $R_{fr}$ increases within this range, the buoyancy nonlinearity intensifies, hindering the inverse transfer of total energy to larger scales. 
The trend with increasing $R_{fr}$ reinforces the results from Section \ref{sub:energy}, where we compared $R_{fr}=1$ (single-phase) and $R_{fr}=\sqrt{2}$.

\begin{figure}%[H]
    \centering
    \includegraphics[width=0.49\textwidth]{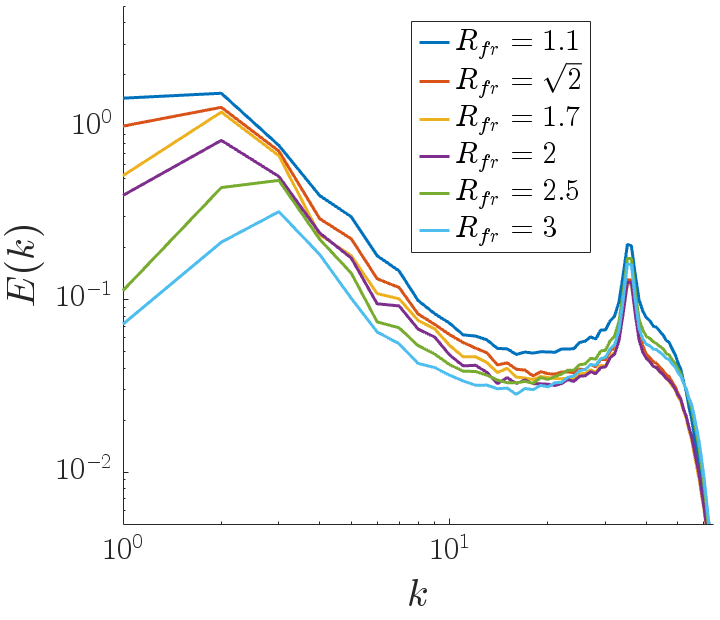}
    \caption{Total energy spectra for different $R_{fr}$ cases at $t=145$.}\label{fig:spectrumNuNs}
\end{figure}

\begin{figure}%[H]
    \centering
    \includegraphics[width=0.49\textwidth]{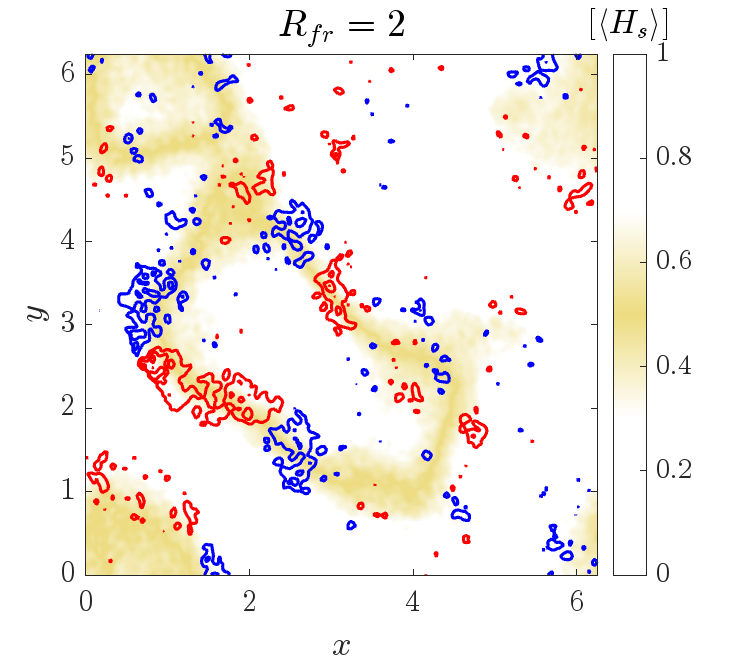}
    \includegraphics[width=0.49\textwidth]{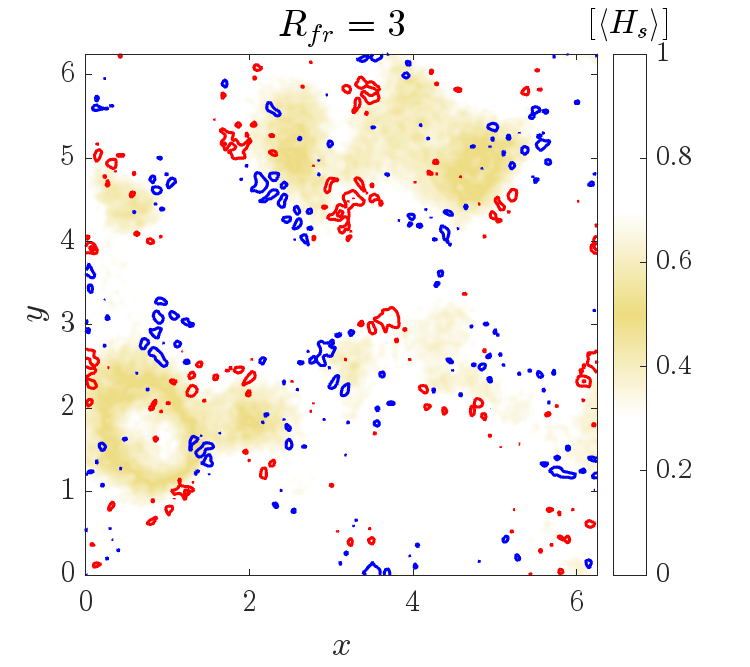}
    
    \caption{Column-integrated, time-averaged $[\langle H_s \rangle ]$ and $[\langle w \rangle ]$ for $R_{fr}=2$ (left) and $R_{fr}=3$ (right). The averaging time window is the same as for the case with $R_{fr} = \sqrt{2}$. The contour levels for $[\langle w \rangle ]$ are chosen as described in Figure \ref{fig:heightave-pv-h-w}.}
    \label{fig:h+w, r_fr=2,3}
    \label{fig:fuzzyboundariesvaryR}
\end{figure}

The physical-space impact of varying $R_{fr}$ is seen by comparing 
Figure \ref{fig:heightave-pv-h-w} (left, $R_{fr} = \sqrt{2}$) to Figure \ref{fig:h+w, r_fr=2,3} (left, $R_{fr}=2$; right, $R_{fr} = 3$).  These figures present the coherent structures corresponding to column-integrated, time-averaged cloud indicator \( [\langle H_s \rangle ]\), and column-integrated, time-averaged vertical velocity 
\( [\langle w \rangle ] \).  Varying $R_{fr}$, one sees structural changes in the yellow regions associated with  high-frequency phase transitions $[\langle H_s \rangle ] \approx 0.5$. Increasing from $R_{fr}=\sqrt{2}$ to $R_{fr} = 2$, thin loop-like areas of yellow become thicker band-like structures and some localized patches.  Increasing further to $R_{fr}=3$, only thick patches of yellow remain.
Concurrently, the structure of \( [\langle w \rangle ]\) also changes, roughly aligned with the structure of $[\langle H_s \rangle ]$. Consistent with decreased energy levels and increased patchiness of phase-transition regions, we also find 
smaller-scale, less-intense $PV_s$ vortices as $R_{fr}$ increases (not shown for conciseness).

Regarding additional wave statistics, it is expected that nonlinear buoyancy leads to non-Gaussian features of waves.  As a signature non-Gaussian feature, we compute the skewness of the time-averaged vertical velocity, defined as 
%conditioned on $\langle H_s \rangle$.  Specifically, we measure
\begin{equation}
{\rm skew}(\langle w \rangle) \equiv \dfrac{\sum_i^N\langle w\rangle^3}{\left(\sum_i^N\langle w\rangle^2\right)^{3/2}}.
\label{eqn:skew}
\end{equation}
Recall that the skewness is identically zero for a Gaussian distribution, and thus we investigate deviation of \eqref{eqn:skew} from zero as a function of the nonlinearity parameter $R_{fr}$. We consider three different samples ($N$): (i) all points in the time-averaged field; (ii) those points in the time-averaged field with $\langle H_s \rangle \in (0.4, 0.6)$ ; and (iii) those points with $\langle H_s \rangle \in (0.45, 0.55)$.  
The sample set (i) yields the global skewness, and sample sets (ii) and (iii) contain only points that undergo frequent change of phase during the time averaging window.
We note that the single-phase case $R_{fr}=1$ with linear buoyancy gives rise to global skewness close to zero, as expected in the dry QG regime.  For phase-change cases
$R_{fr}=1.1, \sqrt{2}, 1.7, 2, 2.5, 3$, Figure \ref{fig:w_skew} (left) shows that the global skewness is also close to zero (blue curve).  The near-zero values can be attributed to two factors.  First, the global data includes the influence of the small-scale random force, which is Gaussian white noise.  Second, it encompasses many data points from regions where \( \langle H_s \rangle \) approaches zero or unity, which are areas without significant phase transitions.
Thus, the global data set is only weakly influenced by the nonlinear nature of waves at phase interfaces.

On the other hand, considering the skewness \eqref{eqn:skew} in regions of high-frequency phase transitions, Figure \ref{fig:w_skew} (left; red and yellow curves), we seen that the skewness of $\langle w \rangle$ decreases as the nonlinearity parameter $R_{fr}$ increases.  In these localized areas, waves continually propagate between liquid and vapor states, and thus the sample points in these areas are maximally influenced by the effects of the nonlinear buoyancy.
The close agreement of the red and yellow curves demonstrates that the deviation trend of local skewness is not sensitive to the cutoff values used to define the regions near $\langle H_s \rangle \approx 0.5$ characterizing high-frequency phase transitions. 
Moreover, whether considering global or local data, the skewness for the case $R_{fr}=1.1$ is nearly zero because there is only a small change in frequency as waves pass through the two different media. As $R_{fr}$ increases, the negative skewness of $\langle w \rangle$ in our setup indicates a preference for downdrafts, and this will be further investigated in future research.

\begin{figure}%[H]
    \centering
    \includegraphics[width=0.49\textwidth]{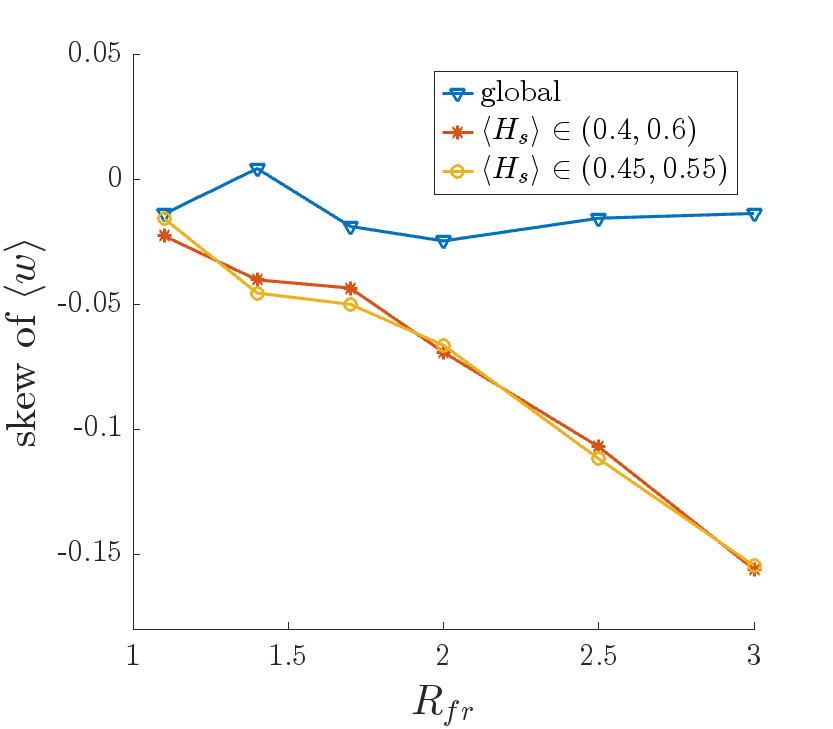}
    \includegraphics[width=0.49\textwidth]{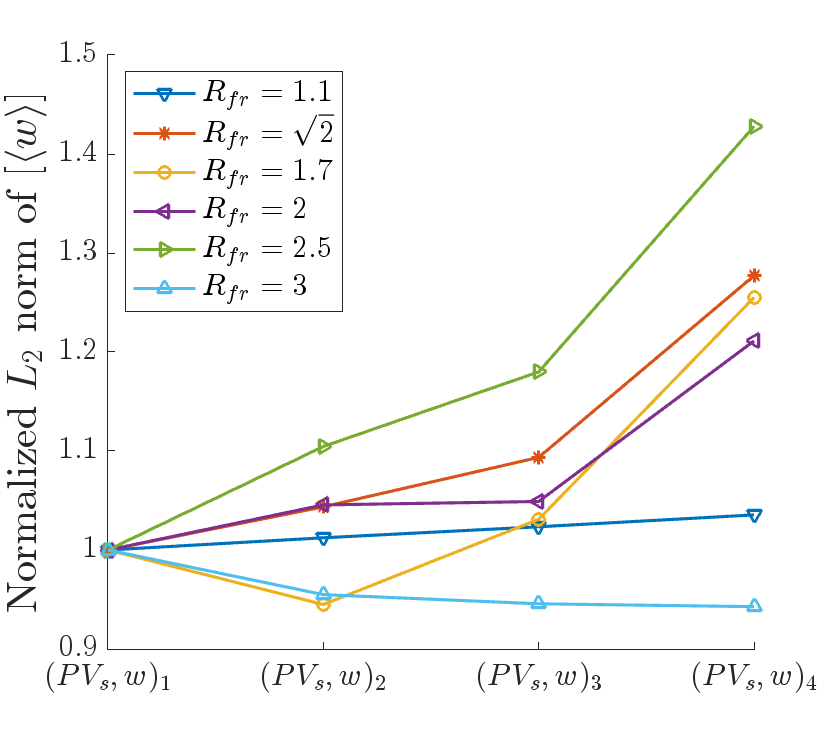}
    
    \caption{Left: skewness \eqref{eqn:skew} of time-averaged vertical velocity versus the ratio of buoyancy frequencies $R_{fr} = N_u/N_s$ for no-rainfall case and the averaging time window is the same as for the case in Table \ref{table: parameter setting}. The blue curve is generated by global data, while datasets for the red and yellow curves are conditioned on high-frequency phase transitions with $0.4 < \langle H_s\rangle < 0.6$ and  $0.45 < \langle H_s\rangle < 0.55$, respectively. Right: normalized $L_2$ norm of $[\langle w \rangle ]$ versus $|[ \langle PV_s \rangle ] |$, where the colors indicate different values of the nonlinearity parameter $R_{fr}$.
    Each curve has four points corresponding to analysis of the filtered datasets $(PV_s, w)_1$ - $(PV_s, w)_4$ as explained in Section \ref{subsec:statwPV_s}; see also Table \ref{table: l2norm-w}.}
    \label{fig:w_skew}
\end{figure}

% To finish the discussion regarding $R_{fr} = N_u/N_s$ dependencies,
To further examine the dependencies of $R_{fr} = N_u/N_f$, we turn to Figure~\ref{fig:w_skew} (right), which is based on the analysis of filtered $(PV_s,w)_f$ datasets described in Section \ref{subsec:statwPV_s}; see also Table \ref{table: l2norm-w}.  The filtering isolates the coherent $PV_s$-vortices together with the associated values of  $[ \langle w \rangle ]$, and then we look for a linkage between high-magnitude values of $[\langle PV_s \rangle ]$ and high-magnitude values of vertical velocity $[\langle w \rangle ]$ (updrafts/downdrafts).  Focusing on the fourth-quartile data points for $\sqrt{2} \leq R_{fr} \leq 2.5$, we see normalized $([ \langle w \rangle ] )_{L_2} \in (1.20,1.45).$ 
%values in the approximate {\color{blue}range $(1.20,1.45).$ what does this mean?  I see from below, use math?}  
The specific values have small dependencies on parameters used in the analysis, such as the exact time-averaging window, depth for the column integration, and the number of standard deviations used to filter out low-level $[\langle PV_s \rangle ]$ data (see Section \ref{subsec:statwPV_s}). On the other hand, upon changing these choices, we found the robust result that fourth-quartile data produces $([\langle w \rangle ])_{L_2} \in  (1.20,1.45)$ for $R_{fr} \in [\sqrt{2},2.5]$.  For this range of $R_{fr}$, we conclude that strong coherent updrafts and downdrafts are associated high-magnitude values of moist potential vorticity.  

However, Figure \ref{fig:w_skew} (right) has two notable outlier cases, namely $R_{fr}=1.1$ (dark blue) and $R_{fr}=3$ (light blue).   For the case $R_{fr}=1.1$, there is essentially no increase in normalized $([\langle w \rangle  ])_{L_2}$ as $| [ \langle PV_s \rangle ]|$ increases.  The lack of increase can be attributed to the weakly nonlinear nature of the waves for $R_{fr} = 1.1$, when the wave frequencies in saturated and unsaturated regions are almost the same.  This means that time averages of waves are small, and therefore induce only weak coupling between $PV_s$ and waves via slow terms in the $PV_s$-evolution equation \eqref{eqn:PV_s-original}.
On the other hand, for the highest value $R_{fr}=3$, when waves are strongly nonlinear, the light blue curve is also flat, which is counter to the argument based solely on nonlinearity of the waves, and seems like a contradiction at first glance.  
The resolution of this seeming contradiction may follow from the fact that the inverse energy transfer is impeded by more frequent phase changes as $R_{fr}$ increases in this range $R_{fr} \in [1.1,3]$.  
Less large-scale energy means overall less dominance by large-scale, coherent structures, and more influence by random, smaller-scale features.  

To conclude this section, we revisit the isolated $(PV_s,w)_f$ datasets. In Table \ref{table: l2norm-w}, the filtered data are divided into four quartiles, which in Figure \ref{fig:l2norm-w} are distinguished by different symbols. For clarity of exposition, however, we simplify the analysis by merging the four quartile subsets into a single region, thereby treating them uniformly. Within this framework, we evaluate the $L_2$ norm of $[\langle w\rangle ]$ in the local $PV$-dominated region relative to its global counterpart. The resulting ratio provides a quantitative measure of the spatial overlap between potential vorticity and vertical velocity. Applying this procedure to both dry and moist simulations, we find that in the moist case with $R_{fr}=\sqrt{2}$ the local-to-global ratio reaches approximately 1.23, whereas the dry runs (case (i) with \(Ro=Fr=0.17\), and case (ii) with \(Ro=0.17,\, Fr=0.24\)) remain close to unity. This demonstrates that phase changes enhance the overlap between $PV_s$ ($PV$ in dry cases) and $w$, thereby intensifying wave–$PV_s$ interactions. Additional parameter regimes, including further $R_{fr}$ cases, are examined in Appendix \ref{subsec:pv-w-overlap}.

% {\color{red}to finish this section..
% refer table 4 and figure 7 considering the isolated dataset PV and W, i.e. (PV, w) set

% to make it simple, wrap them (4-quartile) into to ONE point analysis. i.e in figure 7 we don't distinguish the different symbol, treat them together, uniform one region 

% (global vs local) we look for the L2 of w in local region (PV part) and look for the L2 of w in global region. relative rate between local and global. this relative rate quantitatively indicating the overlap between PV region and W region. the lager relative rate indicating lager overlap

% the same way for dry 1,2

% moist case , more overlap , more linkage, more correlation, more wave-PV interaction.

% Appendix E more case, more $R_fr$
% }

% \begin{color}{red}
% In order to compare the dry and moist runs, we apply the filtering method described in 
% Section \ref{subsec:statwPV_s} to \([\langle PV\rangle]\) in the dry cases (dry case (i) with \(Ro=Fr=0.17\), and dry case (ii) with \(Ro=0.17,\, Fr=0.24\)), measuring the relative size of the 
% \(L_2\) norm of \([\langle w\rangle]\) in the filtered region against its global value; see Appendix \ref{subsec:pv-w-overlap} for details.
% For \(R_{fr}=\sqrt{2}\), the moist run gives about 1.2, while both dry runs remain close to unity. This comparison shows that wave–vortex interaction is stronger in the moist case than in the dry simulations.
% \end{color}

In summary, there is a complicated and delicate balance of effects at play as the ratio $R_{fr} = N_u/N_f$ is increased.
On the one hand, stronger buoyancy nonlinearity (larger $R_{fr})$ may induce stronger coupling between $PV_s$ and $\langle w \rangle$, as suggested by the ODE model of Section \ref{sub-ODE}.  On the other hand, larger $R_{fr}$ inhibits inverse energy transfer and the formation of the large-scale, coherent flow features 
(Figure \ref{fig:spectrumNuNs}). 
\begin{comment}{\color{blue} Additional robustness analysis related to an earlier stopping time in the no-rainfall case is presented in Appendix \ref{subsec:nonprecipvsprecip}.}
\end{comment}

\clearpage
\section{Conclusions}
\label{sec:discussion}

In moist Boussinesq dynamics with water vapor and liquid water, the buoyancy changes its functional form across a phase boundary between unsaturated and saturated environments.  The piecewise buoyancy introduces additional nonlinearity beyond the quadratic nonlinearity, and leads to nonlinear waves (see, e.g., figure 13 of \cite{zhang2021effects}).  Such waves are expected to be dynamically significant in regions of frequent phase changes.  Here we have studied effects of phase transitions and nonlinear waves on inverse energy transfer to large scales and the formation of coherent structures.

We considered the quasi-geostrophic regime with $Fr_s \sim Fr_u \sim Ro = \epsilon < 1$, driven by random forcing at small scales, and in the simplified setting of a triply periodic domain.  In the turbulence literature, this is a classical setting, but here we replace dry, linear buoyancy by nonlinear buoyancy corresponding to warm-rain, bulk cloud physics in the limit of fast cloud microphysics.
The cloud microphysics terms represent condensation of water vapor to form liquid water, and evaporation of liquid water to form water vapor, and they are responsible for the phase transitions and additional nolinearity in the system. 
The investigation of self-organization of random fluctuations is facilitated by the energy source at small scales, which here includes a statistically equal contribution from water vapor and liquid water. 

Compared to dry dynamics, our numerical computations show two main results.  First, frequent phase transitions between vapor and liquid, induced at small scales by random forcing, tend to reduce the inverse transfer of energy.  
In the case without rainfall, inhibition of inverse transfer is attributed to the fact that saturated regions of the flow have higher Froude number $Fr_s$ than unsaturated regions with smaller $Fr_u$ (analogous to the dry Froude number).  Generally speaking, raising the Froude and/or Rossby numbers will slow the inverse transfer to large scales. Rainfall further reduces inverse transfer of kinetic energy by boosting $M$-energy at all scales larger than the forcing scales, thereby decreasing coherence and increasing patchiness of $PV_s$ and averaged wave fields (see Figures \ref{fig:energy_1}, \ref{fig:M-energy}, \ref{fig:heightave-pv-h-w}, 
\ref{fig:heightave-h-w}).

Second, and most importantly, the energy that does accumulate at large scales has special features that are distinct from dry dynamics. 
In dry dynamics, the slowly varying potential vorticity is weakly coupled to waves because waves are linear at lowest order, and thus their time-averages tend to zero at lowest order.   By contrast, the moist potential vorticity $PV_s$ is more strongly coupled to non-zero time averages of nonlinear waves.
Since buoyancy non-linearity appears in regions of frequent phase transitions, wave-$PV_s$ coupling leads to an inextricable linkage between large-scale, fuzzy phase boundaries, coherent (up)downdrafts, and $PV_s$-vortices (Figures \ref{fig:heightave-pv-h-w}, \ref{fig:h+w, r_fr=2,3}).
%(Figures \ref{fig:heightave-pv-h-w}, \ref{fig:h+w, r_fr=2,3}, \ref{fig:heightave-h-w}).  
As shown quantitatively in Section 4.6, the wave–PV interactions in the moist case at $Ro \approx  0.017$ is about 19 times stronger than in the dry case. Altogether, the morphology of the large-scale coherent structures is determined by complex dynamics introduced by the buoyancy nonlinearity parameter $R_{fr}=Fr_s/Fr_u$.
The ODE model of Section \ref{sub-ODE} suggests that increasing $R_{fr}$ strengthens wave-$PV_s$ coupling. However, Figure \ref{fig:spectrumNuNs} shows that that increasing $R_{fr}$ decreases the inverse transfer to large scales. A balance between such competing effects ultimately determines the thickness of high-frequency phase-transition regions, the size and strength of $PV_s$-vortices, as well as the locations and intensity of the (up)downdrafts. 
%(Figures \ref{fig:heightave-pv-h-w}, \ref{fig:h+w, r_fr=2,3}). 

\begin{comment}
As mentioned, the presence of rain with fall-speed $V_r=1$ impedes inverse transfer by increasing $M$-energy at scales larger than the forcing scales, leading to broader regions of frequent phase changes and less-energetic coherent structures.
\end{comment}

%(Figure \ref{fig:heightave-h-w}).
%partial fragmentation of the coherent structures. 

The current study helps to characterize inverse energy transfer and the formation of coherent structures for moist Boussinesq dynamics in the presence of phase changes.  There remain open mathematical problems to pursue, such as rigorous fast-wave-averaging analysis and mathematical
representation of the free-boundary dynamics. With applications in mind, it would be interesting to examine other parameter regimes, more realistic boundary conditions, and targeted geophysical scenarios including additional physical effects.  For example, future work will address potential vorticity dipoles observed in moist convective storm tracks with $Ro = O(1)$ and wind shear \citep{chagnongray09,weijenborgetal2015,weijenborgetal2017}.

\section*{Acknowledgements}
LMS thanks Annick Pouquet for encouraging completion of this study.  LMS and YZ have benefited from the insights of Antoine Remond-Tiedrez and Sam Stechmann over many years leading up to the research presented herein.  
YZ was partially supported by National Natural Science Foundation of China grants 12241103, 12401562 and Shanghai Pujiang Program grant 22PJ1403500. LMS gratefully acknowledges support by the National Science Foundation, Division of Mathematical Sciences DMS-1907667, as well as Deutsche Forschungsgemeinschaft (DFG) through the Research Unit FOR5528.   The authors thank three anonymous reviewers whose insightful comments helped to improve the manuscript.

\section*{Declaration of interests} The authors report no conflict of interest.

\clearpage

\appendix

\section{Selected simulation results for $Ro \approx 0.017$, $Fr_u \approx 0.017$ and $Fr_s \approx 0.024$}
\label{subsec: smaller parameters}

Here we present the analog of Figure \ref{fig:heightave-pv-h-w} for smaller $Ro \approx 0.017$, $Fr_u \approx 0.017$ and $Fr_s \approx 0.024$ (Figure \ref{fig: heightave-pv-h-w, appA}). The top row of the figure is data from the simulation of moist dynamics with phase changes ($V_r=0$), and the bottom row is data from the dry simulation.  Comparing Figures \ref{fig:heightave-pv-h-w} and \ref{fig: heightave-pv-h-w, appA}, one sees that the overall conclusions presented in the manuscript are consistent for a reduction in the non-dimensional parameters, moving closer to the quasi-geostrophic limit.  In particular, there is a tight linkage between time-averaged quantities $\langle H_s\rangle$, $\langle w \rangle$ and $\langle PV_s \rangle$ in moist dynamics. Furthermore, in the moist dynamics with phase changes, the time-averaged waves $\langle w \rangle$ are dynamically more important for the formation of large-scale coherent updrafts and downdrafts.  This is attributed to the nonlinear buoyancy, leading to nonlinear waves near phase interfaces. In the dry case, however, only a single buoyancy exists. Therefore, the nonlinear interaction associated with phase change is absent, and the vertical velocity $\langle w\rangle$ is strongly reduced, with coherent updrafts and downdrafts nearly vanishing.

As discussed in Section \ref{subsec:dynamicwaves}, we quantitatively examine the nonlinear interactions, namely wave-$PV_s$ in the moist and wave-$PV$ in the dry case, and find clear differences between the two simulations. For $Ro=0.17$, the moist interaction is about 2 times stronger than the dry case, while for $Ro=0.017$ the difference increases to nearly 19 times.

% {\color{red}quantitatively speaking, as discussed in section 4.6, when mentioned the nonlinear interaction (specifically wave-$PV_s$ in moist case while wave-PV in dry ), we find in $Ro=0.17$ case , difference between dry and moist, 2 times stronger. in $Ro=0.017$ case , 19 times stronger.}

% {\color{blue}
% In the phase-change case ($Ro = Fr_u = 0.017, Fr_s = 0.024$), we found that $R_\textrm{avg}/R_0 \approx 0.664$, which is nearly 19 times larger than the value of 0.035 obtained in the dry simulation ($Ro = Fr = 0.017$).
% }

\begin{figure}
\centering
\begin{minipage}{0.49\textwidth}
\includegraphics[width = 0.98\textwidth]
{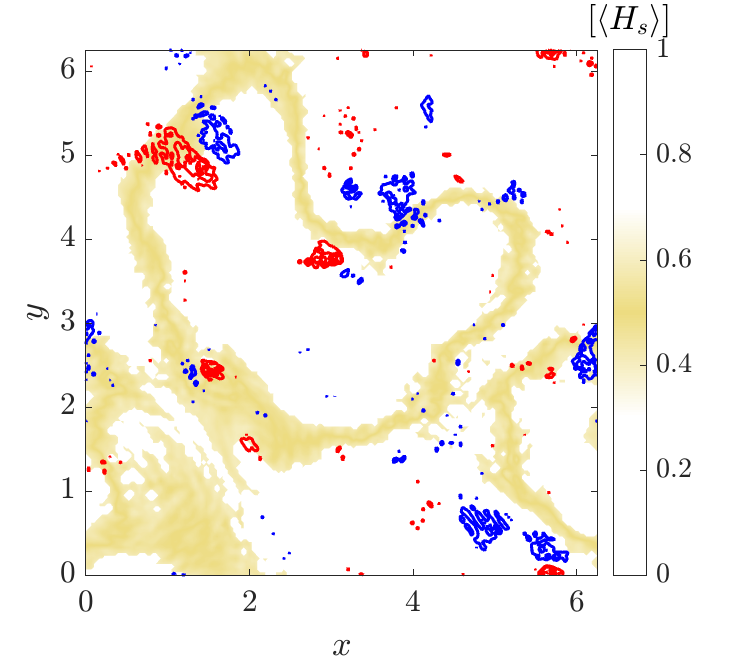}
\end{minipage}
\begin{minipage}{0.49\textwidth}
\includegraphics[width = 0.98\textwidth]
{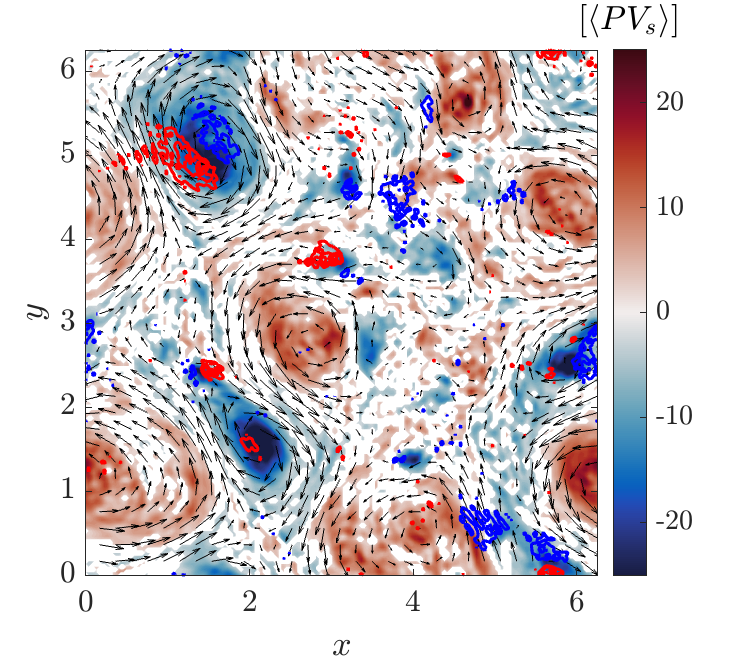}
\end{minipage}
\begin{minipage}{0.49\textwidth}
\includegraphics[width = 0.98\textwidth]
{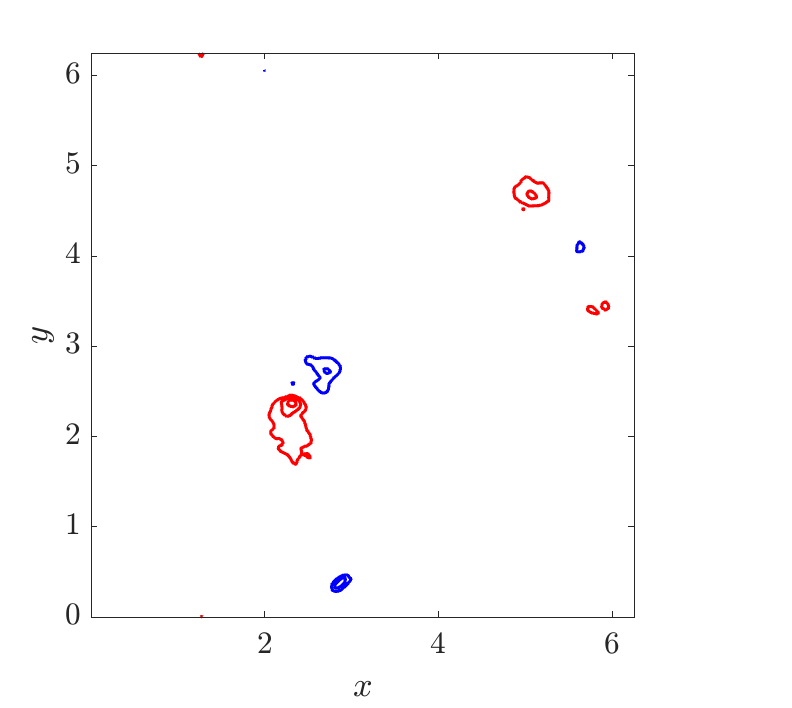}
\end{minipage}
\begin{minipage}{0.49\textwidth}
\includegraphics[width = 0.98\textwidth]
{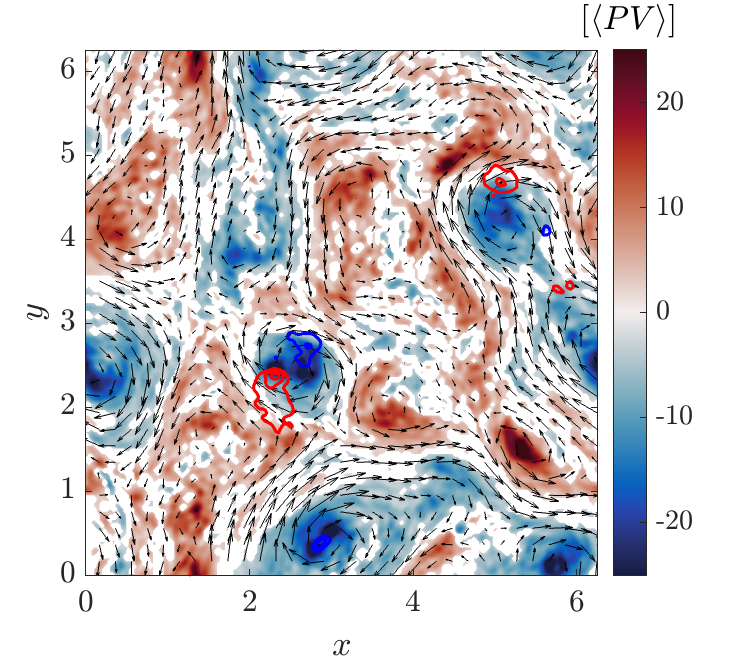}
\end{minipage}
\caption{The analog of Figure \ref{fig:heightave-pv-h-w} for $Ro \approx 0.17$, $Fr_u \approx 0.017$ and $Fr_s \approx 0.024$.  The peak forcing wavenumber is $k_f = 36$ and the resolution is $N=192$. All other information is the same as in Figure \ref{fig:heightave-pv-h-w}.}
\label{fig: heightave-pv-h-w, appA}
\end{figure}

\section{Structure of precipitating flow in the periodic domain}
\label{subsec:nonprecipvsprecip}

Figure \ref{fig:heightave-h-w} is an analog of Figure \ref{fig:heightave-pv-h-w}, but now for precipitating flow with $V_r=1$.  The left panel displays $[\langle w \rangle ]$ and $[ \langle H_s \rangle ]$ for averaging window beginning at time \( t = 145 \). 
Compared to nonprecipitating flow with $V_r=0$, the entire structure is significantly more fragmented
and replete with small-scale features, consistent with lower energy at large scales as seen in Figure \ref{fig:energy_1} (right).
Areas of high-frequency phase-transitions occupy a much larger fraction of the total area. Updrafts and downdrafts are scattered across the domain, although their rough association with phase boundaries is still evident.  Simultaneously, the $PV_s$-vortices are smaller, less coherent and less intense 
(Figure \ref{fig:heightave-h-w} right).
We recall from Section \ref{subsec:M-energy} that rainfall boosts the $M$-energy associated with energy exchange between unsaturated and saturated potential energies, i.e., associated with phase transitions (Figure \ref{fig:M-energy}). The relative contribution of total potential energy increases compared to kinetic energy (Tables \ref{table3} and \ref{table4}), reducing inverse transfer of kinetic energy, and consequently reducing the size and strength of the $PV_s$-vortices.

\begin{figure}%[H]
    \centering
    \includegraphics[width=.49\textwidth]{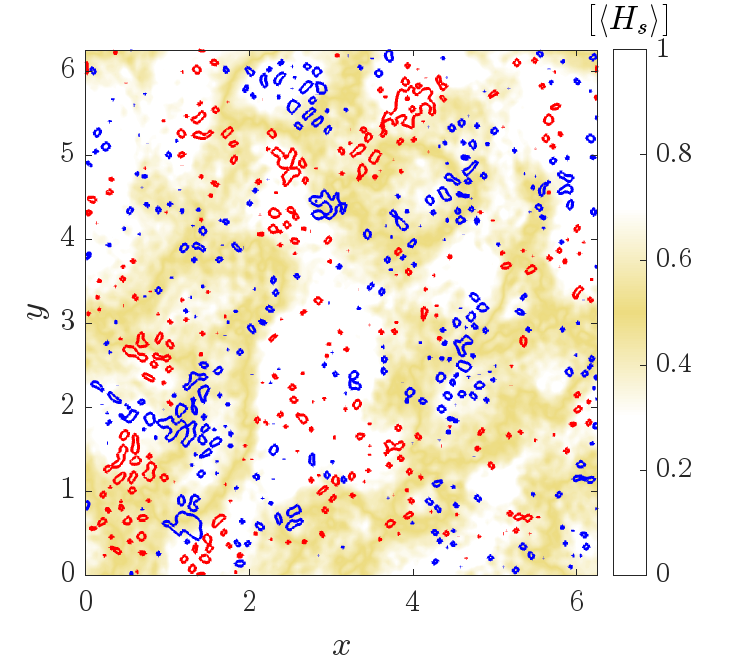}
    \includegraphics[width=.49\textwidth]{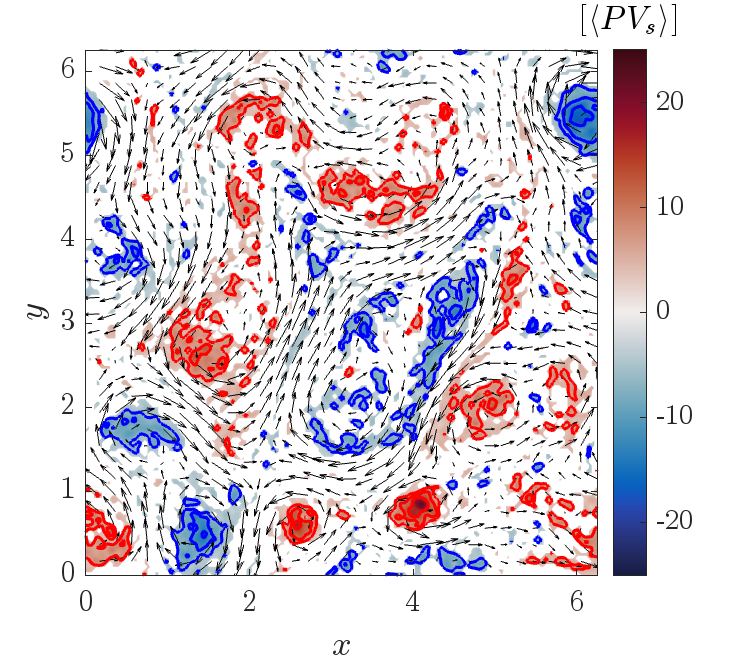}
    \caption{Flow structure for the rainfall case with $V_r=1$ for comparison to Figure \ref{fig:heightave-pv-h-w}.
 Left: Column-integrated, time-averaged $[\langle H_s\rangle]$ with contours of vertical velocity $[\langle w\rangle]$ (red for positive and blue for negative).  Right: $[\langle PV_s\rangle ]$ with horizontal velocity vectors $[\langle {\bf u}_h \rangle ] = [\langle u\rangle] \hat{\bf x}$ + $[\langle v\rangle ] \hat{\bf y}$ (black arrows).  See Figure \ref{fig:heightave-pv-h-w} for more details. }
 \label{fig:heightave-h-w}
\end{figure}

\section{3D visualization of coherent structures}
\label{subsec:3D_HwPV_s}

In Section~\ref{sub-column}, we explored column-integrated, time-averaged structures, highlighting the connection between phase interfaces $[\langle H_s \rangle ]$, vertical velocity \( [\langle w \rangle ]\), and moist potential vorticity \( [\langle PV_s \rangle ] \) in the absence of rainfall (\( V_r = 0 \)). %The 2D visualizations provide insight into the spatial relationships of these quantities, showing that
%they are all closely linked together. 
Here we zoom in on two representative regions of interest, identified on the 2D plots (the boxes in the top right panel of Figure~\ref{fig:heightave-pv-h-w}), to illustrate representative 3D structures in the time-averaged fields.  
%The larger sub-box 1 suggests a downdraft in between counter-rotating $PV_s$-vortices, while the smaller sub-box 2 indicates updrafts and downdrafts on opposite sides of an anticyclonic $PV_s$-vortex.

\begin{figure}
    \centering
\includegraphics[width=0.49\textwidth]
{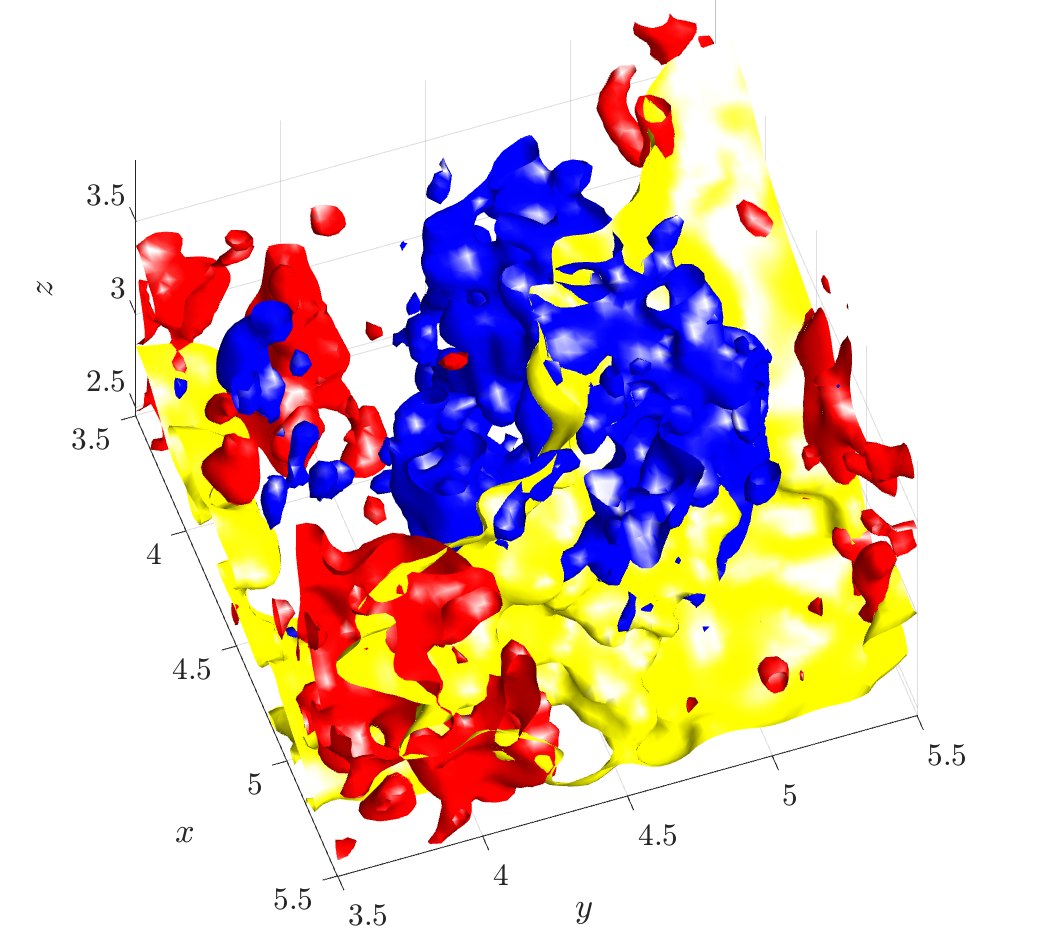}
\includegraphics[width=0.49\textwidth]
{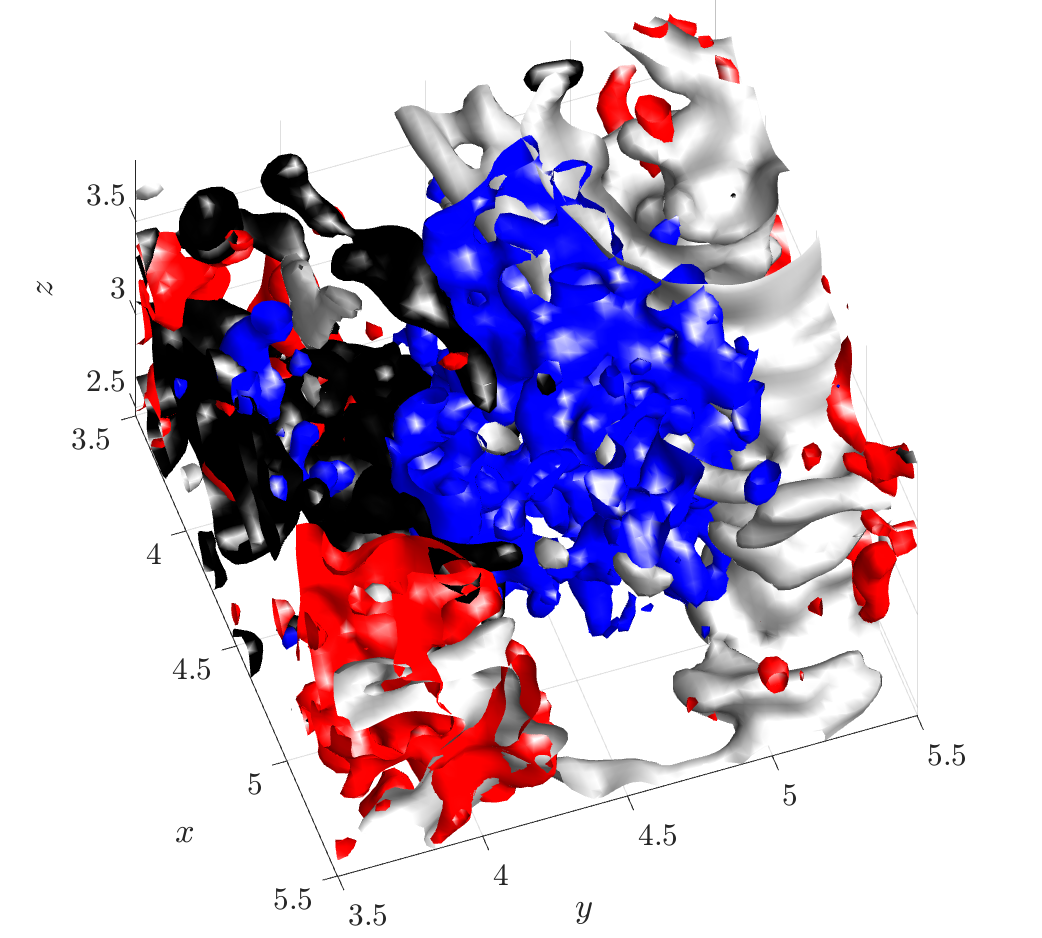}
\caption{Isosurfaces of $\langle H_s\rangle$, $\langle w \rangle$ and 
$\langle PV_s\rangle$ for the large sub-box 1 in the upper right panel of 
Figure~\ref{fig:heightave-pv-h-w}.
Left: The yellow phase boundary has value $\langle H_s\rangle = 0.5$; red indicates an isosurface inside the updraft with value $\langle w \rangle = 0.25 \; \rm{max}(|\langle w \rangle|) = 0.17$; the blue indicates an isosurface inside the downdraft with 
$\langle w \rangle = -0.17$.
Right: The red and blue isosurfaces of $\langle w \rangle$ are the same as on the left; grey (black) is an isosurface of positive (negative) $\langle PV_s \rangle$ with value $\langle PV_s \rangle = 0.25 \; \max(|\langle PV_s\rangle|) = 20$ (-20).}
    \label{fig:3d-h-w}
\end{figure}

\begin{figure}
\centering
\includegraphics[width=0.49\textwidth]
{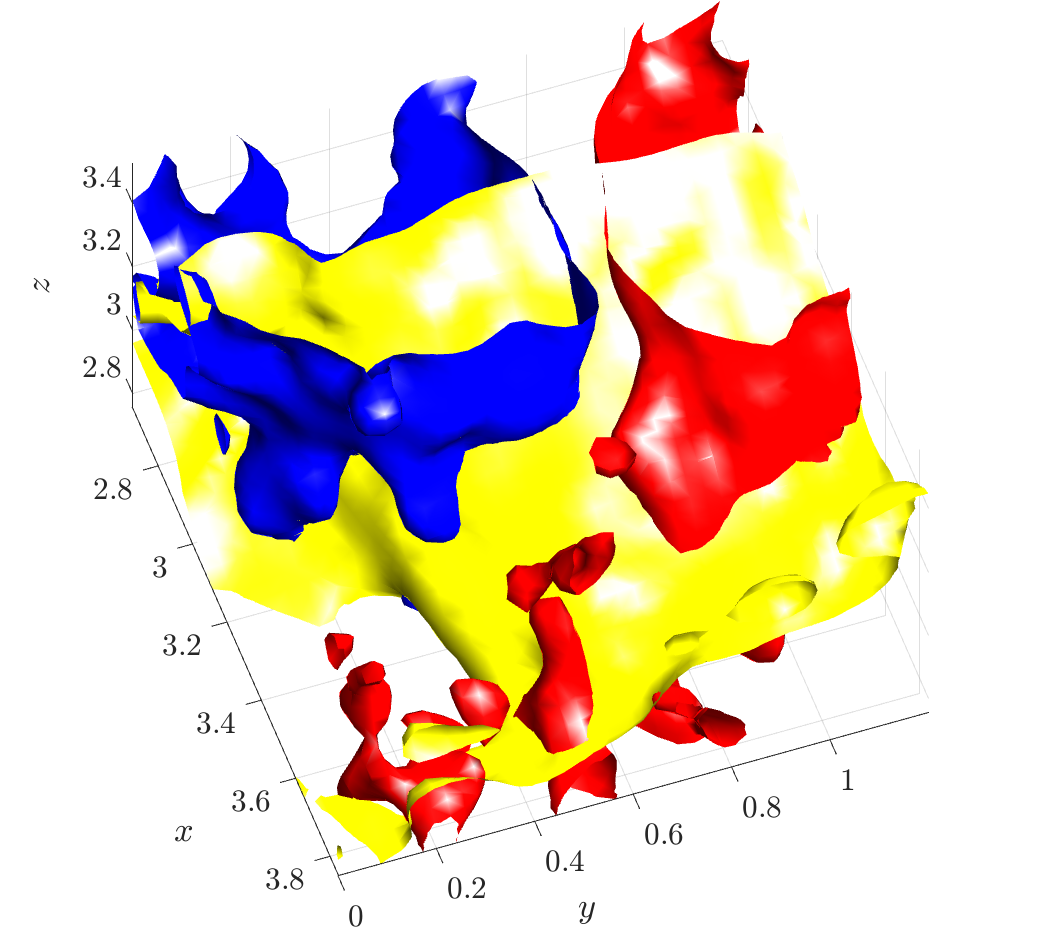}
\includegraphics[width=0.49\textwidth]
{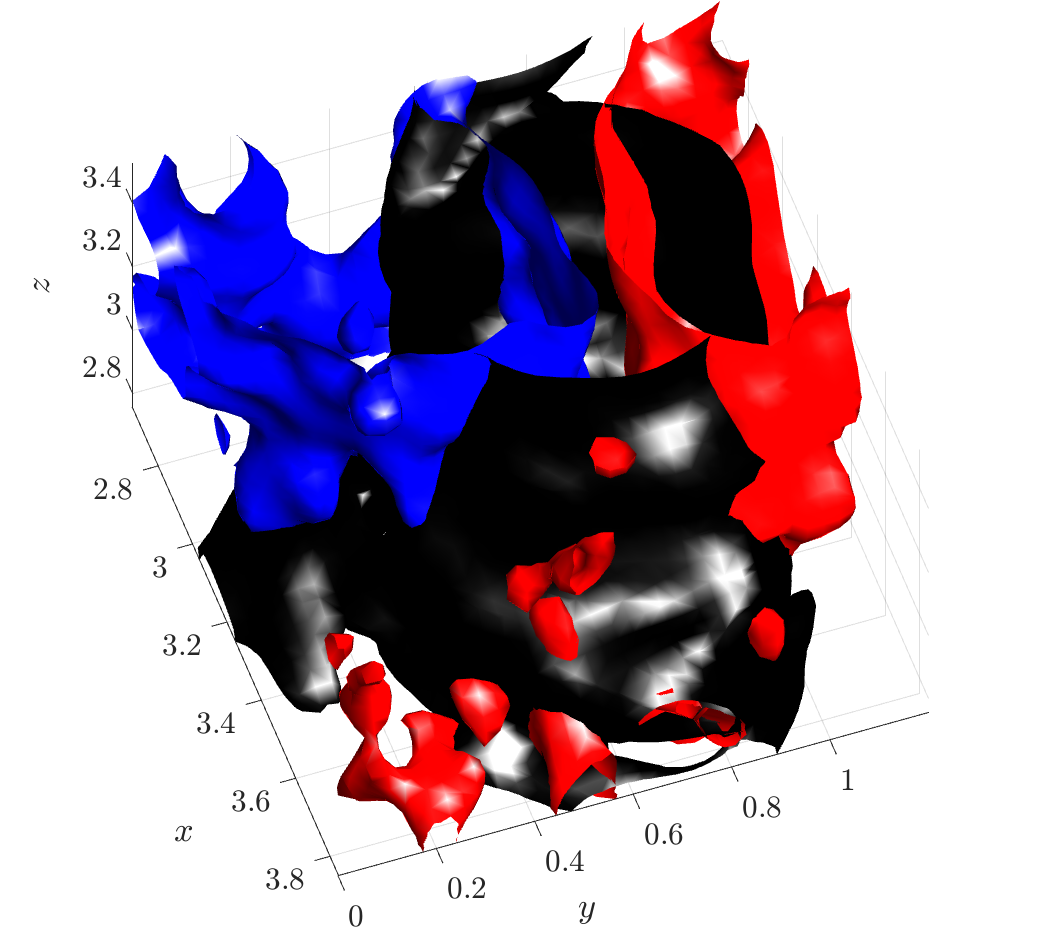}
\caption{Isosurfaces of $\langle H_s\rangle$, $\langle w \rangle$ and 
$\langle PV_s\rangle$ for the small sub-box 2 in the upper right panel of 
Figure~\ref{fig:heightave-pv-h-w}. The levels are identical to those in Figure \ref{fig:3d-h-w}.
}
\label{fig:3d-pv}
\end{figure}

Figures \ref{fig:3d-h-w} and \ref{fig:3d-pv} show isosurfaces of $\langle H \rangle$, $\langle w \rangle$ and 
$\langle PV_s \rangle$ for, respectively, sub-box 1 and sub-box 2.  On the left panel of both figures, one can see that the yellow phase interface with isosurface value $\langle H_s \rangle =0.5$ cuts through the isosurfaces of $\langle w \rangle$, where red (blue) indicates an isosurface of the updraft (downdraft) (see the figures for detailed information).  For sub-box 1, the right panel of Figure \ref{fig:3d-h-w} confirms a downdraft region horizontally in-between positive (cyclonic) and negative (anticyclonic) $PV_s$-structures.
There is also an updraft horizontally adjacent to, and vertically below, negative $PV_s$.
Corresponding to sub-box 2, and as anticipated, the right panel of 
Figure \ref{fig:3d-pv} shows an anticyclonic region of negative $PV_s$ flanked by an updraft and a downdraft. 

In this disordered flow forced randomly at small scales, 
Figures~\ref{fig:3d-h-w}-\ref{fig:3d-pv} show variability of the patterns associated with the slowly-evolving large-scale flow, and yet these figures also re-iterate two common features.  First, coherent updrafts and downdrafts appear prominently in regions of frequent phase changes, and second, these updrafts and downdrafts are in close proximity to $PV_s$-vortices.

\section{Slow-fast decomposition and time-averaged flow fields}
\label{subsec:slow-fast-coupling}

As mentioned in Section \ref{subsec:fast-slow-variables}, \cite{ss17} showed that $PV_s$ and $M$ given by \eqref{def:PV_s}-\eqref{def:M-1} are slowly varying quantities that characterize the moist dynamics of coherent structures \citep[see also][] {wssm19,wssmz20}. The quantities $PV_s$ and $M$  are referred to as slow because 
they are in the kernel of the leading-order (nonlinear) operator, and they do not change in time in the limit $Fr_s \sim Fr_u \sim Ro = \epsilon \rightarrow 0$.

In flow with nonlinear buoyancy \eqref{eqn:buoyancy}, the variables $(u,v,\theta_e,q_t)$ can be separated into two distinct contributions. The first contribution is directly related to slow variables $PV_s$ and $M$ by the definitions \eqref{eqn:bu-bs-def}, \eqref{def:PV_s} and \eqref{def:M-1}, and is therefore denoted by 
$(u_\textrm{(pv,m)}, v_\textrm{(pv,m)},{\theta_e}_\textrm{(pv,m)}, {q_t}_\textrm{(pv,m)})$.   The second contribution is the wave part, such that a scalar flow field $f$ may be expressed as 
\begin{equation}
f_\textrm{(wave)} = f - f_\textrm{(pv,m)} .
\label{def:wave}
\end{equation} 
Note that one can also define the waves 
%component 
explicitly as the image of the leading-order operator
\citep{remmel2009new,hss14,remondtiedrez-decomposition2024}.
Note also that the vertical velocity $w$ is a purely wave quantity with $w_\textrm{(pv,m)}=0$ (as in dry dynamics).
For \( Fr_s \sim Fr_u \sim Ro = \epsilon \ll 0 \) and time-averaging windows $\tau \gtrsim 1$, the time average of a mixed quantity can be computed from \eqref{def:wave}:
\begin{equation}
\langle f \rangle = 
\langle f_{\textrm{(pv,m)}} \rangle + \langle f_{\textrm{(wave)}} \rangle
\approx f_{\textrm{(pv,m)}} + \langle f_{\textrm{(wave)}} \rangle.
\label{def:slowwave}
\end{equation}

The procedure for finding $(u_\textrm{(pv,m)}, v_\textrm{(pv,m)},{\theta_e}_\textrm{(pv,m)}, {q_t}_\textrm{(pv,m)})$ follows from the definitions for $PV_s$ and $M$ together with the `balance relations' (geostrophic and hydrostatic balance) that appear at $O(\epsilon^{-1})$ in the asymptotic expansion of \eqref{eqn:boussinesq} as $\epsilon \rightarrow 0$ \citep{ss17,remondtiedrez-decomposition2024}.  This procedure has two steps, here collectively referred to as $PV_s$-and-$M$ inversion \citep{ss17}: (i) given $PV_s$ and $M$, a nonlinear elliptic equation determines a slow pressure $\psi$, and then (ii) $\psi$ and $M$ define $({{\bf u}_h}_\textrm{(pv,m)},{\theta_e}_\textrm{(pv,m)}, {q_t}_\textrm{(pv,m)})$ according to the balance relations and variable definitions for $\epsilon \rightarrow 0$.  The non-dimensional version of $PV_s$-and-$M$ inversion is given by 
\begin{align}
&\nabla_h^2\psi + {\cal F}^2 \frac{\partial^2 \psi}{\partial z^2} + \dfrac{\partial}{\partial z} \biggl [ \dfrac{1}{1+{\cal G}}\text{min}({\cal F}M - {\cal F}^2{\cal G}\frac{\partial \psi}{\partial z}, 0) \biggr ] = PV_s,
\label{PVsMinversion}
\\
&(u_\textrm{(pv,m)},v_\textrm{(pv,m)}) =(-\frac{\partial \psi}{\partial y}, \frac{\partial \psi}{\partial x}) \quad\textrm{geostrophic  balance},
\label{balance-1}\\ &{\theta_e}_\textrm{(pv,m)} = {\cal F}\frac{\partial \psi}{\partial z} + \dfrac{1}{1+{\cal G}} \textrm{min}\biggl(M- {\cal F}{\cal G}\frac{\partial \psi}{\partial z},0 \biggr)   \quad \textrm{hydrostatic balance}, 
\label{balance-2}\\
& {q_t}_\textrm{(pv,m)} = M - {\cal G}{\theta_e}_\textrm{(pv,m)}\\
%\quad \textrm{definition of} \; M, \label{balance-3}\\
%& {b_s}_\textrm{(pv,m)} = {\theta_e}_\textrm{(pv,m)} - C_{cl} Ro \cdot{q_t}_\textrm{(pv,m)} 
& {b_s}_\textrm{(pv,m)} = {\theta_e}_\textrm{(pv,m)}\\  
%\quad \textrm{definition of} \; b_s \; \textrm{for} \; \epsilon \rightarrow 0,\\
& {b_u}_\textrm{(pv,m)} = {\theta_e}_\textrm{(pv,m)}  - {q_t}_\textrm{(pv,m)}
\quad \; %\textrm{definition of} \;
%b_u \; \textrm{for} \; \epsilon \rightarrow 0,
 \label{balance-relations}
\end{align}
where ${\cal F} = Fr_s/Ro$, and ${\cal G} = 
Fr_s (Fr^{-2}_u - Fr^{-2}_s)^{1/2}$. 
%and $C_{cl}= c_p \theta_0/L_v.$
%Similarly, one can extract ${b_u}_\textrm{(pv,m)}$ from its definition in terms of $\theta_e$ and $q_t$, but we omit this formula for brevity.

Altogether, \eqref{def:wave}-\eqref{balance-relations} provide a way to separate vortical and wave contributions of flow quantities in the regime $Fr_s \sim Fr_u \sim Ro = \epsilon \ll 1$.  This separation allows us to assess their relative contributions to the evolution of $PV_s$ (and $M$).

 \section{
 % Comparison of $PV–w$ overlap under different parameters
Comparison of wave–potential vorticity interactions under different parameters
}
\label{subsec:pv-w-overlap}

This appendix quantifies the relation between potential vorticity and vertical velocity across parameter ranges in both moist cases ($PV_s$) and dry cases ($PV$).
Following the analysis framework as mentioned in Section \ref{subsec:NuNsimpact}, we evaluate the $L_2$ norm of $[\langle w\rangle]$ inside a local $PV$-dominated region and compare it with its global counterpart.
The local region is defined by a $[\langle PV_s\rangle]$-based filter
\begin{equation}
\{\,PV_s \mid PV_s \notin (\mu \pm a\sigma)\,\},
\label{pv-filter}
\end{equation}
where $\mu$ and $\sigma$ denote the mean and standard deviation of $[\langle PV_s\rangle]$ as discussed in Section~\ref{subsub:region_selection_sta}. 
% Here we choose  $a=2$ to characterize the vortex region. 
% The reported ratio is the local $L_2$ norm divided by the global $L_2$ norm, providing a concise quantitative measure of the spatial overlap between potential vorticity and vertical velocity. A larger value implies stronger correspondence.
The reported ratio of the local to global $L_2$ norm serves as a quantitative measure for the degree of spatial overlap between potential vorticity and vertical velocity, where a larger value suggests a stronger correspondence.

% The results in Table~\ref{tab:local_global_ratio} show that for the moist cases with 
% $R_{fr}=1.1, \sqrt{2}, 1.7, 2,$ and $2.5$, the local-to-global ratios lie in the range 
% 1.19--1.29, whereas the dry runs remain close to unity (0.93--1.02). 
% This corresponds to an enhancement of roughly 20\% in the moist regime, 
% indicating that phase changes substantially strengthen wave-$PV$ interactions.

Results in Table~\ref{tab:local_global_ratio} indicate that for moist cases ($R_{fr}=1.1, \sqrt{2}, 1.7, 2, 2.5$), the local-to-global ratios range between 1.19--1.29, in contrast to the dry runs, which remain near unity (0.93--1.02). This approximate 20\% enhancement in the moist regime suggests that phase changes may play a substantial role in strengthening wave-$PV_s$ interactions.

\begin{table}
\centering

\begin{tabular}{l|cccccccc}
\hline
$R_{fr}$ & 1.1 & $\sqrt{2}$ & 1.7 & 2 & 2.5 & 3 & dry case (i) & dry case (ii) \\ %& proportion \\
% \hline
% global norm& 0.045 & 0.053 & 0.047 & 0.050 & 0.045 & 0.046 & 0.045 & 0.047 &  \\
\hline
% $a=1$ & & & & & & & & & 31.7\% \\
% local norm & 0.049 & 0.057 & 0.048 & 0.053 & 0.048 & 0.046 & 0.045 & 0.046 & \\
% ratio & 1.085 & 1.081 & 1.025 & 1.060 & 1.074 & 1.000 & 0.985 & 0.985 & \\
% \hline
% $a=1.1$ & & & & & & & & & 27.1\% \\
% local norm & 0.049 & 0.058 & 0.049 & 0.054 & 0.049 & 0.046 & 0.045 & 0.046 & \\
% ratio & 1.096 & 1.094 & 1.032 & 1.074 & 1.083 & 1.002 & 0.987 & 0.989 & \\
% \hline
% $a=1.3$ & & & & & & & & & 19.4\% \\
% local norm & 0.050 & 0.059 & 0.049 & 0.056 & 0.050 & 0.046 & 0.045 & 0.047 & \\
% ratio & 1.119 & 1.117 & 1.047 & 1.105 & 1.112 & 1.015 & 0.985 & 0.991 & \\
% \hline
% $a=1.5$ & & & & & & & & & 13.4\% \\
% local norm & 0.051 & 0.061 & 0.050 & 0.058 & 0.052 & 0.047 & 0.044 & 0.047 & \\
% ratio & 1.139 & 1.147 & 1.070 & 1.145 & 1.150 & 1.020 & 0.980 & 0.991 & \\
% \hline
% $a=1.7$ & & & & & & & & & 8.9\% \\
% local norm & 0.052 & 0.063 & 0.053 & 0.058 & 0.054 & 0.046 & 0.044 & 0.046 &  \\
% ratio & 1.168 & 1.177 & 1.121 & 1.161 & 1.201 & 1.002 & 0.976 & 0.985 &  \\
% \hline
% $a=2$ & & & & & & & &  \\ %& 4.6\% \\
% local norm & 0.053 & 0.066 & 0.057 & 0.061 & 0.058 & 0.046 & 0.042 & 0.048 & \\
ratio & 1.195 & 1.237 & 1.208 & 1.207 & 1.290 & 1.000 & 0.934 & 1.023 \\%& \\
\hline
% $a=2.5$ & & & & & & & & & 1.2\%\\
% local norm & 0.053 & 0.072 & 0.066 & 0.067 & 0.065 & 0.045 & 0.036 & 0.049 & \\
% ratio & 1.188 & 1.347 & 1.391 & 1.330 & 1.460 & 0.0987 & 0.790 & 1.045 & \\
% \hline
% $a=3$ & & & & & & & & & 0.3\% \\
% local norm & 0.052 & 0.081 & 0.073 & 0.069 & 0.070 & 0.049 & 0.033 & 0.048 & \\
% ratio & 1.172 & 1.520 & 1.544 & 1.376 & 1.569 & 1.070 & 0.733 & 1.028 & \\
% \hline
\end{tabular}
\caption{Local-to-global $L_2$-norm ratios of $[\langle w\rangle]$ for moist cases at different $R_{fr}$ and for two representative dry cases (dry case (i) with $Fr = Ro = 0.17$, and dry case (ii) with $Fr = 0.24$, $Ro = 0.17$). }
\label{tab:local_global_ratio}

\end{table}

\clearpage

% \bibliographystyle{jfm}
% \bibliography{main-jfm-ref, antoine-references}
\bibliographystyle{abbrvnat}
\bibliography{main-jfm-ref,antoine-references}
% \bibliography{antoine-references}

\end{document}